\documentclass[]{interact}
\usepackage[colorlinks=true, linkcolor=blue, citecolor=blue, urlcolor=blue]{hyperref}
\usepackage{booktabs}  
\usepackage{array}     
\usepackage{tabularx}  
\usepackage{float}

\usepackage{ulem}

\usepackage{color}

\usepackage{placeins}
\usepackage{epstopdf}
\usepackage[caption=false]{subfig}
\usepackage[numbers,sort&compress]{natbib}

\def\NAT@def@citea{\def\@citea{\NAT@separator}}
\makeatother

\theoremstyle{plain}

\theoremstyle{definition}

\theoremstyle{remark}

\begin{document}

\articletype{Review Article}

\title{Advances in dual-chirped optical parametric amplification}

\author{
\name{Kaito Nishimiya and Eiji J. Takahashi\thanks{CONTACT Eiji J. Takahashi Email: ejtak@riken.jp RIKEN Center for Advanced Photonics, RIKEN, 2-1 Hirosawa, Wako, Saitama 351-0198, Japan}}
\affil{RIKEN Center for Advanced Photonics, RIKEN, 2-1 Hirosawa, Wako, Saitama 351-0198, Japan}
}

\maketitle

\begin{abstract}
High-energy infrared lasers have enabled the generation of strong field phenomena, and among such phenomena, high-order harmonic generation (HHG) from gases has enabled attosecond-scale observations in atoms or molecules.
Lasers with longer wavelengths and shorter pulse widths are advantageous for generating higher photon energy and shorter attosecond pulses via HHG. Thus, the development of ultrashort mid-infrared (MIR) lasers has progressed.
This paper reviews research on developing high-energy MIR lasers using the dual-chirped optical parametric amplification (DC-OPA) method. 
We developed TW-class multi-cycle lasers in the MIR region, which was previously difficult. 
The advanced DC-OPA method, an extension of the conventional DC-OPA method, enables one-octave amplification of the wavelength, and a TW-class single-cycle laser was developed.
These lasers were utilized for HHG, enabling single-shot absorption spectroscopy, and one-octave supercontinuum soft X-ray generation for single-cycle isolated attosecond pulse.
We also show the development of multi-TW sub-cycle DC-OPA pumped by Ti:sapphire laser and high average power MIR single-cycle DC-OPA using thin-disk laser technology. 
\end{abstract}


\section{Introduction}
The advent of high-energy ultrashort lasers has enabled realization of relativistic intensity levels ($>$10$^{18}\,$W/cm$^2$). Previous studies investigated phenomena associated with strong laser fields, e.g., high energy electron/ion beam generation through laser acceleration \cite{LWFA1,LWFA2,LWFA3}, ultrafast X-ray generation via surface plasma oscillations \cite{SHHG1,SHHG2}, and attosecond pulse generation through high-order harmonic generation (HHG) in gases \cite{HHG1,HHG2,HHG3,HHG0321_1,HHG_321_2} and so on \cite{strongphy}. 
HHG has been applied to ultrafast research to track and control the motion of electrons in atoms or molecules \cite{atto_electron1,atto_electron2,atto_electron3,atto_electron4,atto_electron5}. 
Laser pulse energy and the number of laser cycles within a pulse envelope are key parameters to characterize strong laser field phenomena. 
For example, reducing the number of laser cycles, which corresponds to shortening the laser pulse width, can yield phenomena not observed in multi-cycle lasers, e.g., enhancing ion beam acceleration efficiency \cite{IBA_single} and isolated attosecond pulse (IAP) generation \cite{IAP1,IAP2,IAP3}, and so on\cite{IAP4}.
Especially, The number of laser cycles in the pulse envelope is considered the most crucial parameter in HHG research.



\begin{figure}
\centering\includegraphics[width=8cm]{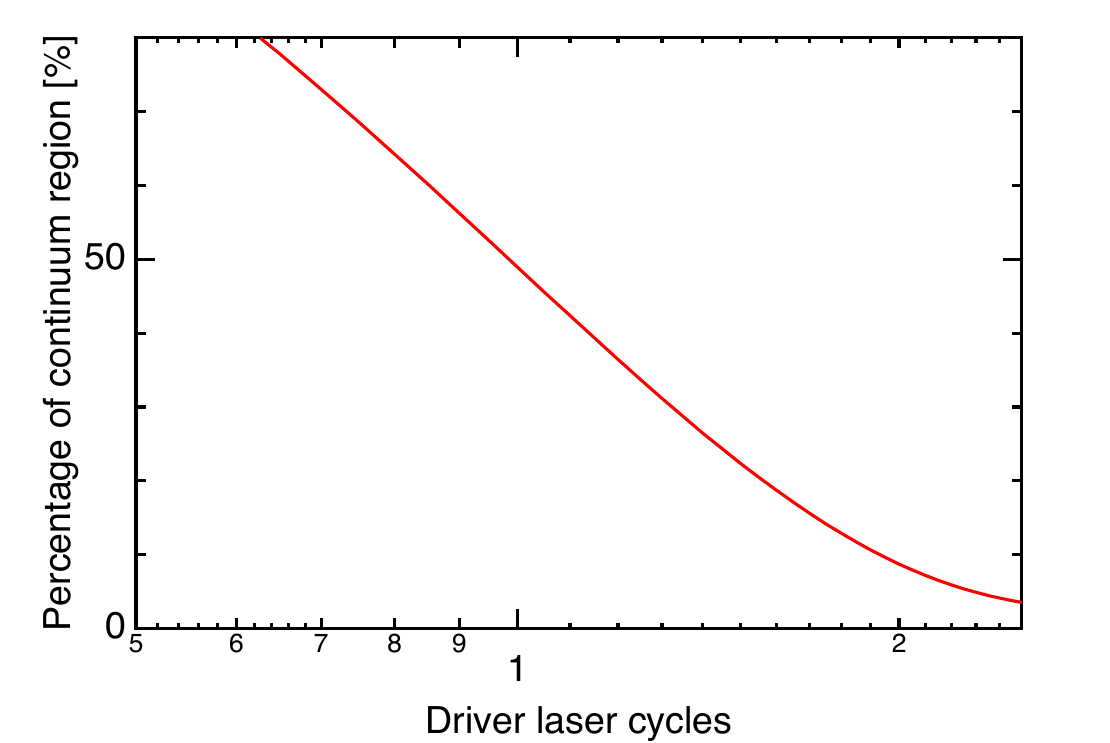}
\caption{Relationship between cycle number of the fundamental laser and percentage of the HH spectrum continuum region.} 
\label{cycyle vs cont}
\end{figure}

Attosecond pulse generation via HHG is based on the framework of the semiclassical so-called three-step model, and the maximum photon energy of HHG is approximately given by $E_{\rm cutoff} {\rm [eV]} $ = $I_p$ + 3.17 $U_p$ (cutoff low) \cite{CO_rule1,CO_rule2}, where $I_p$ denotes the binding energy of the electrons, and $U_p {\rm [eV]} = 9.38 \times 10^{-14} I {\rm [W/cm^2]} (\lambda {\rm [\mu m])}^2$ is the electron quiver energy (i.e., the ponderomotive energy).
Thus, research on generating attosecond pulses with higher photon energy and shorter pulse width is generally synonymous with driver laser development. 
Previously, the driver laser was based on a Ti:sapphire crystal \cite{TiS} and chirped pulse amplification (CPA) \cite{CPA1,CPA2}. 
In a typical CPA system, a Ti:sapphire laser has a pulse width of over several tens of fs and a laser cycle of 10 or more (unless it is shaped spectrally) \cite{TiS10fs1,TiS10fs2,TiS10fs3}. 
A high-order harmonic (HH) beam is generated when electrons recombine every half cycle of the driver laser \cite{3step}; thus, when a multi-cycle laser is employed as a driver laser, the HHs become an attosecond pulse train (APT) \cite{rabbit,APTAC}. 
On the other hands, an IAP does not have a repeat temporal structure and is preferable to an APT because it is well-suited for ultrafast applications.
To create an IAP via HHG, the continuum region (near the cutoff energy of the HH) must be extracted using a filter or multilayer mirror. 
Figure \ref{cycyle vs cont} shows the relationship between the number of driver laser cycles and the percentage of continuum regions in the entire HH spectrum.
Note that reducing the number of laser cycles expands the continuum regions in the HH spectrum, which enables generation of shorter IAPs.

Previous studies have developed post-compression techniques for a few-cycle pulse generation \cite{HCF1,HCF2} and carrier envelop phase (CEP) stabilization technology \cite{CEP1,CEP2}, and IAP generation has been demonstrated \cite{IAP1,IAP2,IAP3}.
Polarization gating \cite{attochirp2,PG2,PG3}, amplitude gating \cite{PhysRevLett.104.233901,syn2}, and double optical gating \cite{DOG1,DOG2} can generate an intense IAP using multi-cycle lasers. 
The IAPs generated by Ti:sapphire lasers have been employed to develop techniques to observe the motion of electrons \cite{atto_electron1,atto_electron2,atto_electron3,atto_electron4,atto_electron5}, IAP pulse width measurements \cite{FROG_CRAB}, and nonlinear extreme ultraviolet optics \cite{RN955,RN947,NonlOpt1,NonlOpt2}. 
However, the available photon energy of IAP generated by a Ti:sapphire laser in real-world applications has been approximately below 100 eV due to its low photon flux \cite{67as}.

To expand attosecond applications, both the photon flux and photon energy are important factors. 
According to the cutoff low, the maximum photon energy of the HH can be expressed as: $E_{\rm cutoff}\propto I\lambda^{2}$. 
A Ti:sapphire laser system had been a promising driver laser for HHG; however, the driver laser intensity $I$ must be increased beyond the ionization threshold of the target medium to generate a few hundred eV HH \cite{TiSWW1,TiSWW2}.
Conversion efficiency of HHG is extremely drop due to the fully ionized state of the target medium; thus, increasing the cutoff energy efficiently by increasing the driver laser intensity $I$ is difficult.
Methods to realize efficient HHG from the ionized media include quasi-phase matching \cite{QPM1,QPM2,QPM3,QPM4} and nonadiabatic self-phase matching \cite{NSPM}; however, these methods are difficult to control and are impractical. 
It is important to suppress the ionization of the medium during HHG and satisfy the phase matching (PM) condition in a neutral medium to improve the conversion efficiency \cite{PM}. 
From the perspective of the pulse energy scaling of IAPs, the combination of loose focus technology \cite{loose1,loose2} and satisfying the PM conditions in neutral gases \cite{MIRPM} is a promising approach.
Thus, ultrafast MIR lasers are ideal driver lasers for higher photon energy IAP generation because the cutoff energy is proportional to the square of the driver laser wavelength $\lambda$.
However, 
the conversion efficiency decreases in proportion to the -5th to -6th power of the driver laser wavelength \cite{CO_rule3}. 
Therefore, the pulse energy of the MIR laser is extremely important to compensate for the loss of conversion efficiency. 
 A previous study that considers ultrafast MIR driver laser reported an HHG experiment with a maximum photon energy exceeding 1 keV \cite{keV}.

\begin{figure}[t]
\centering\includegraphics[width=0.8\textwidth]{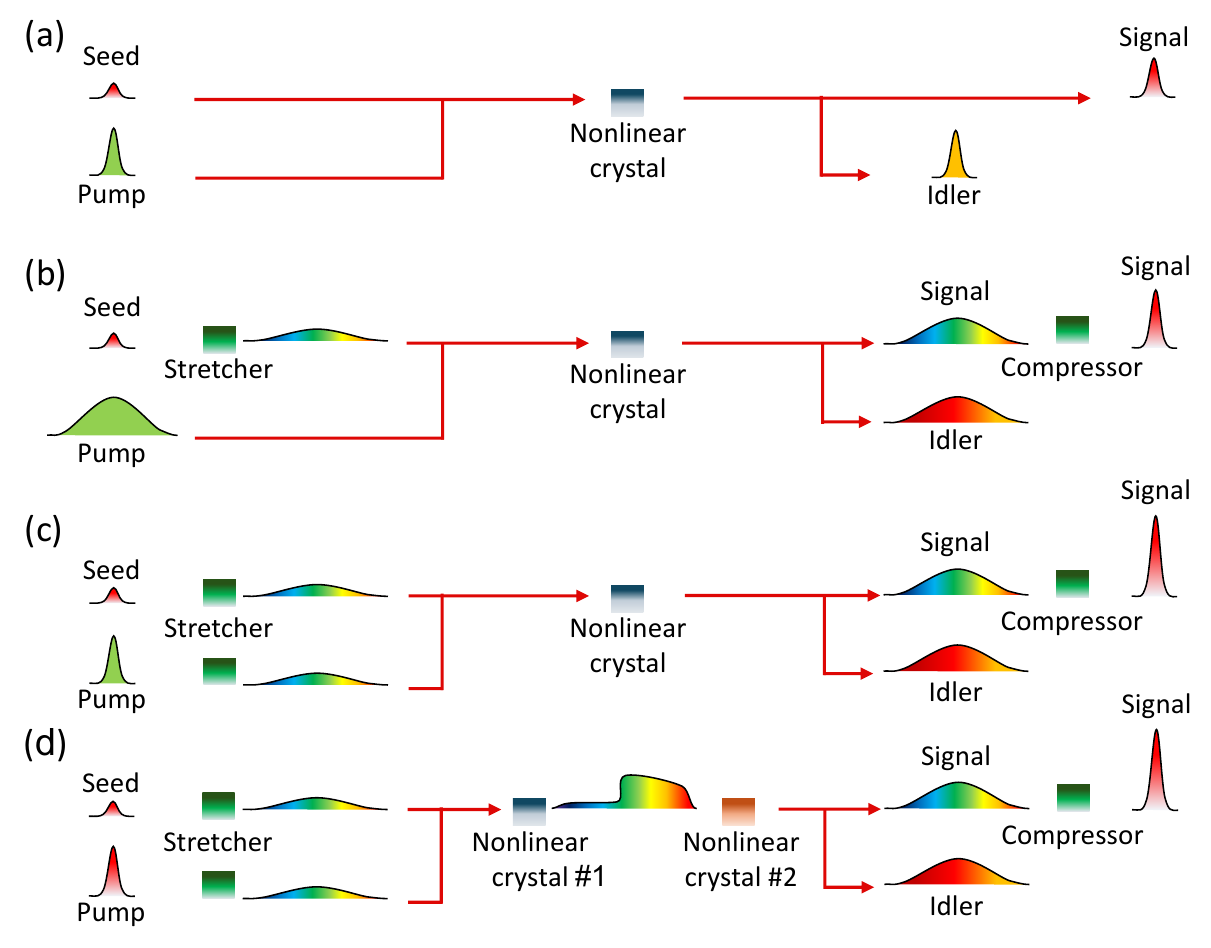}
\caption{Concepts of (a) OPA, (b) OPCPA, (c) DC-OPA, and (d) advanced DC-OPA with heterogeneous nonlinear crystals (\#1, \#2).} \label{OPA concept}
\end{figure}

Ultrafast MIR laser have been actively developed, primarily through optical parametric amplification (OPA) (Fig. \ref{OPA concept}(a)) \cite{OPA,OPA1,OPA2}.
Many MIR lasers that utilize OPA have been developed \cite{OPA3,OPA4,OPA5,PostCompref65,OPA7,OPA8,OPA9}; however, their pulse energies are commonly limited to approximately 1 mJ due to damage to the nonlinear crystal. 
Ross $et~al.$ proposed optical parametric chirped pulse amplification (OPCPA) (Fig. \ref{OPA concept}(b)) \cite{OPCPA} to overcome the poor energy scalability of OPA. 
OPCPA reduces the intensity by chirping the seed in time and utilizing a picosecond laser for the pump to avoid damaging the crystal, similar to CPA. 
OPCPA has enabled laser development with higher pulse energy than conventional OPA \cite{OPCPA,OPCPA1,OPCPA2,OPCPA3,OPCPA4}: however, the pump and the seed should be synchronized temporally. 
There are some few-cycle lasers in MIR region with OPCPA; however, their energy has been limited to the mJ level \cite{OPCPAMIR}. 
The frequency domain OPA (FOPA) method \cite{FOPA1,FOPA2} was proposed as a sub-two-cycle amplification method, and it has achieved 30 mJ and 11 fs; however, it is difficult to stabilize the CEP (due to spatial dispersion) and to obtain higher pulse energy.
Other methods to generate few-cycle MIR pulses include the optical waveform synthesizer \cite{syn1,syn3}, post-compression, filamentation \cite{sub_Post1,sub_post2}, and OPA with a special pump wavelength \cite{subopa}; however, these methods are limited to pulse energies of approximately 1 mJ.

We proposed the dual-chirped optical parametric amplification (DC-OPA) method in 2011 (Fig. \ref{OPA concept}(c))\cite{DCOPA_cal}, which chirps both the pump and the seed, unlike OPCPA, which only chirps the seed. 
Using a broadband Ti:sapphire laser for the pump, the amplified bandwidth in OPA can be expanded, and higher pulse energy can be obtained efficiently compared with conventional OPCPA.
Here, the pump and the seed are created from a single-laser system; thus, temporal synchronization can be achieved easily, and the wavelength can be tuned by selecting a nonlinear crystal appropriately. 
It has been demonstrated that the advanced DC-OPA method, which utilizes heterogeneous nonlinear crystals, can amplify a single-cycle pulse \cite{xu_single}. 
This method employs two kinds of nonlinear crystals for amplification, making it possible to amplify a bandwidth of one octave or more, which was difficult (Fig. \ref{OPA concept}(d)) with the conventional DC-OPA method.

This paper discusses the development of TW-class MIR laser using the DC-OPA method and its application to attosecond pulse generation.
This paper is organized as follows.
Section \ref{TW-class MIR DC-OPA} describes the development of a TW-class MIR laser system using the DC-OPA method. 
Section \ref{TW-class MIR single-cycle laser based on advanced DC-OPA} discusses the development of a TW-class single-cycle laser system with an advanced DC-OPA method.
In Section \ref{Future prospect}, future prospects of the DC-OPA method are explained.
Finally, the paper is concluded in Section \ref{Conclusion}.

\section{TW-class MIR DC-OPA}
\label{TW-class MIR DC-OPA}

\subsection{Features of DC-OPA}
\label{Features of DC-OPA}

\begin{figure}[h]
\centering\includegraphics[width=0.7\textwidth]{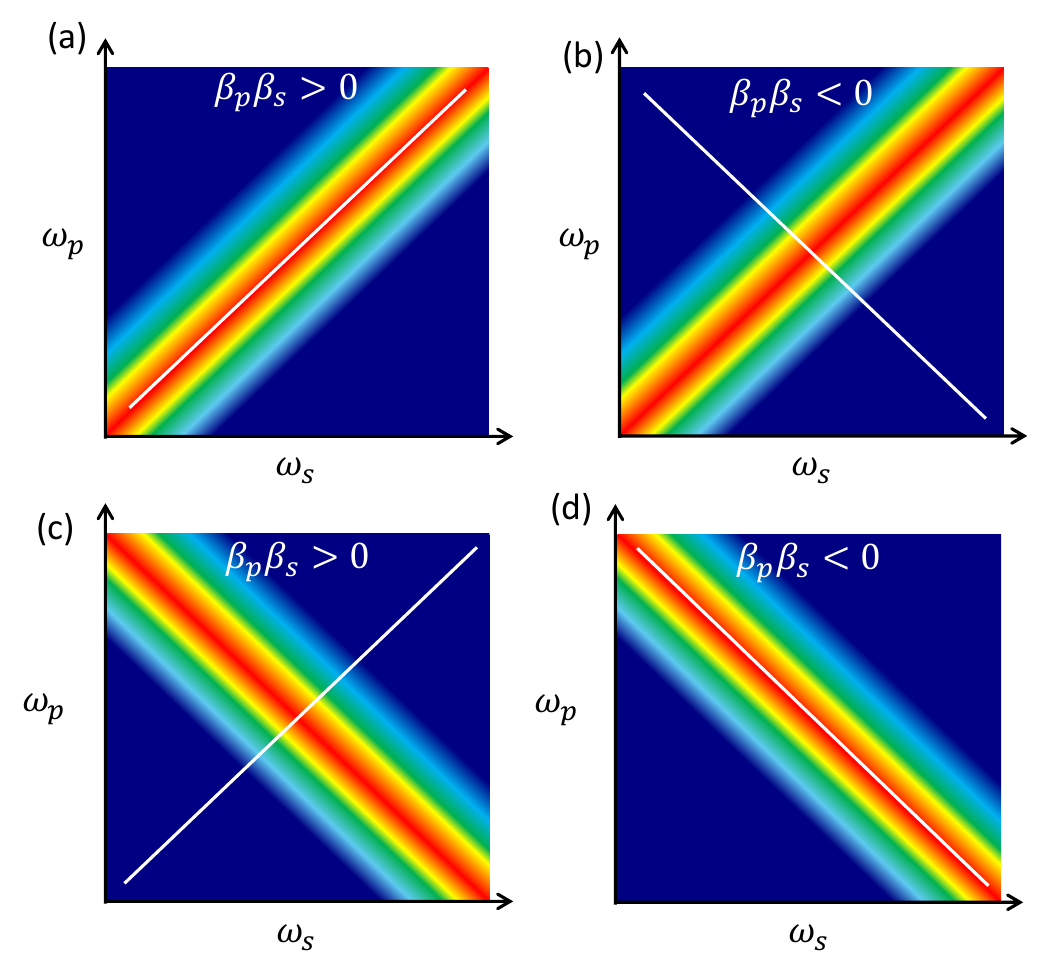}
\caption{Relationship between chirp matching and PM. The horizontal and vertical axis represent the angular frequency of the signal(seed) and the pump. PM is satisfied with an upward slope with (a) $\beta_{\rm{p}} \beta_{\rm{s}}>0$ and (b) $\beta_{\rm{p}} \beta_{\rm{s}}<0$. (c) PM is satisfied with a downward slope with$\beta_{\rm{p}} \beta_{\rm{s}}>0$ and (d)$\beta_{\rm{p}} \beta_{\rm{s}}<0$.} \label{chirp match}
\end{figure}

DC-OPA is a flexible amplification method which is similar to OPA. By changing the optical axis and the kind of amplifying nonlinear crystal, amplification at various wavelength region can be realized, even with the same pump wavelength \cite{DCOPA_cal}.
PM conditions and chirp matching are important to realize the flexibility of the DC-OPA method. 
First, as with conventional OPAs, the amount of phase mismatch $\Delta k$ is an important parameter. 
The conversion efficiency of an OPA is proportional to $\mathrm{sinc}(\Delta kL/2)^2$, where $L$ denotes the medium length; thus, a value where $\Delta k$ is close to 0 must be used. 
In the DC-OPA method, the pump and the signal are chirped; therefore, the PM condition must be satisfied at the wavelength where the respective components coincide in time, i.e., chirp matching. 
Here, the pump, signal and idler angular frequency in the DC-OPA are denoted $\omega_{\rm{p}}$, $\omega_{\rm{s}}$, and $\omega_{\rm{i}}$, respectively; then: 
\[
 \omega_ p =  \omega_s+ \omega_i
\]
When the pump and signal have a linear chirp $\beta_{\rm{p}} $ and $\beta_{\rm{s}} $, then the angular frequencies at a certain time $t$ are expressed as:
\[ \omega_ {\rm{p}}(t) =  \omega_{\rm{p0}}+\beta_{\rm{p0}} (t+\Delta t) \]
\[ \omega_ {\rm{s}}(t) =  \omega_{\rm{s0}}+\beta_{\rm{s}} t \]
where $\omega_{\rm{p0}}$ and $\omega_{\rm{s0}}$ denote the central angular frequency of the pump and the signal, respectively, and $\Delta t$ denotes the delay time between the pump and the signal, which can be adjusted arbitrarily. 
Eliminating $t$ from these two equations, we obtain: 
\[ \omega_{\rm{p}} =  \dfrac{\beta_{\rm{p}}}{\beta_{\rm{s}}}\omega_{\rm{s}}+\omega_{\rm{p0}} - \dfrac{\beta_{\rm{p}}}{\beta_{\rm{s}}}\omega_{\rm{s0}}+\beta_{\rm{p}}\Delta t\]
In other words, the relationship between $\omega_{\rm{p}} $ and $\omega_{\rm{s}} $ depends on the amount of chirp ($\beta_{\rm{p}} $ and $\beta_{\rm{s}} $), and chirp matching is achieved by adjusting these values.
A conceptual diagram of PM and chirp matching is presented in Fig. \ref{chirp match}. 
In Fig. \ref{chirp match} (a), the PM condition is upward-sloping; thus, chirp matching must also be upward-sloping, i.e., the pump and signal chirp signs must match. 
When chirp matching is not achieved (Fig. \ref{chirp match}(b)), the wavelength region in which PM is satisfied becomes narrow; thus, the amplification bandwidth becomes narrow, and the conversion efficiency become low. 
If the PM slope is in the opposite direction (Fig. \ref{chirp match}(c,d)), chirp matching can be satisfied by reversing the chirp direction of the pump and signal (Fig. \ref{chirp match}(d)), which corresponds to reversing the time sequence of the pulse.
Note that chirp matching also affects the idler, i.e., the pump minus the signal, which affects time, frequency, and space. 
For example, if the signal and the pump chirp amounts are equal, the idler will be chirp-free \cite{Yin:18}, and if they are arranged noncollinearly, the idler has angular dispersion \cite{DCOPA4}.
Chirp matching is an important parameter when amplifying a broad bandwidth with DC-OPA. 
In this calculation, a linear chirp is assumed for the pump and the signal; however, with broadband amplification, PM and chirp matching should consider the signal's nonlinear chirp.

\subsection{DC-OPA configuration and demonstration}
\begin{figure}[th]
\centering\includegraphics[width=0.8\textwidth]{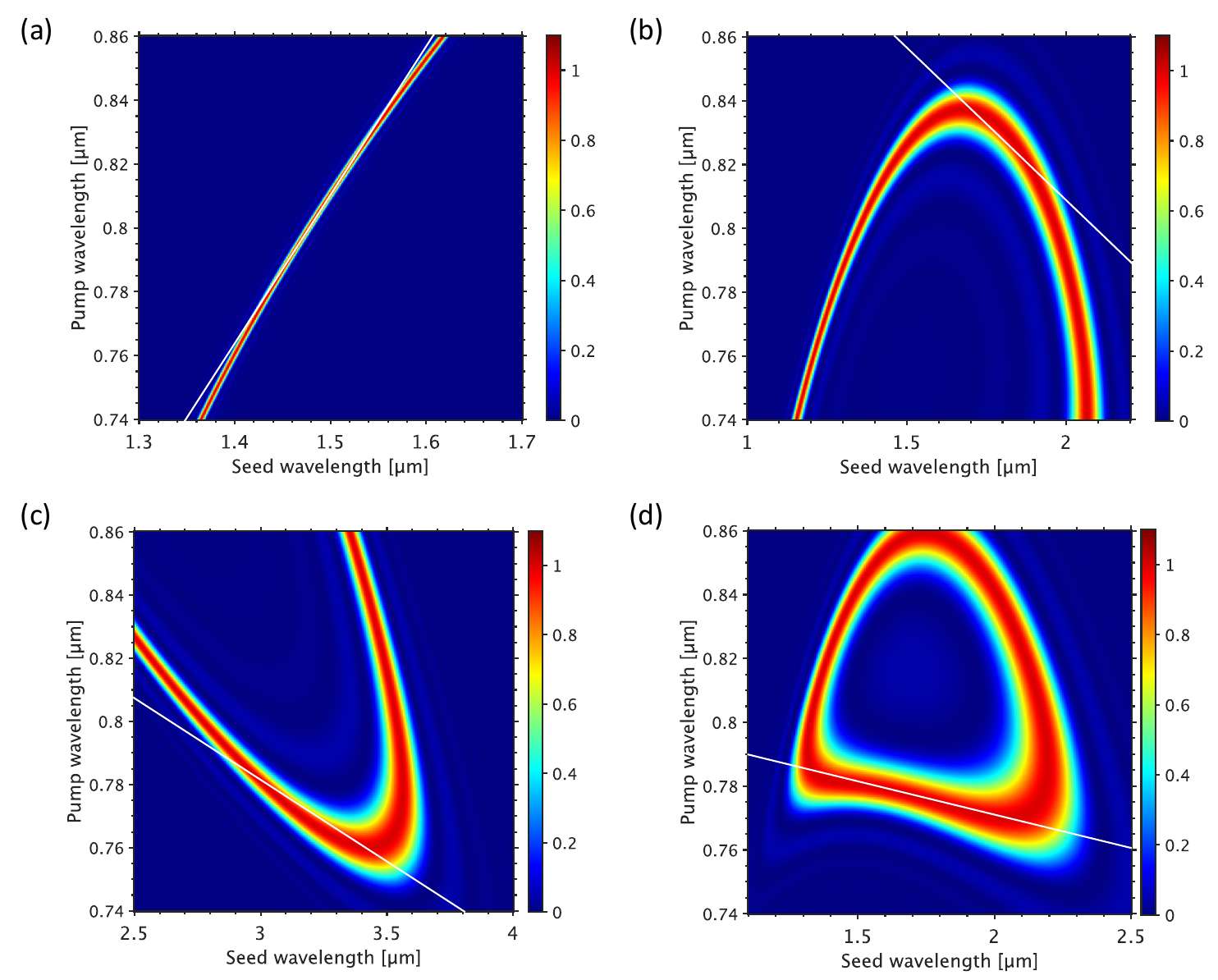}
\caption{PM of (a) type II BBO ($\theta=29.5^{\circ}$), (b) type I BBO ($\theta=20.2^{\circ}$), (c) type I MgO:LN ($\theta=48.75^{\circ}$), and (d) type I BiBO ($\theta=10.9^{\circ}$) crystals. The white lines show the chirp matching assuming a linear chirp of the pump and the seed.} \label{phase}
\end{figure}

To realize efficient amplification over the broadest possible bandwidth in each crystal, we first calculated the PM conditions and assembled an experimental device that could realize chirp matching in a broad bandwidth.
Several PM and chirp matching conditions used in our experiment are plotted in Fig. \ref{phase}, which shows the PM conditions of type I $\beta$-$\rm{BaB_2O_4}$ (BBO), type II BBO, type I 5 mol$\%$ MgO-doped $\rm{LiNbO_3}$ (MgO:LN), and type I $\rm{BiB_3O_3}$ (BiBO) crystals. 
Using 750--850 nm of the Ti:sapphire gain spectrum as the pump and carefully designing the optical axis of the nonlinear crystals, it is possible to amplify a wavelength region of 1.3--2$\,\mu$m with the type I \cite{DCOPA4} and type II BBO crystals \cite{1.7umDCOPA}, 2--4 $\,\mu$m with the type I MgO:LN crystal \cite{3umDCOPA}, and 1.2--2.5$\,\mu$m with the BiBO crystal \cite{sub2DCOPA}. 

\FloatBarrier

\begin{figure}
\centering\includegraphics[width=0.8\textwidth]{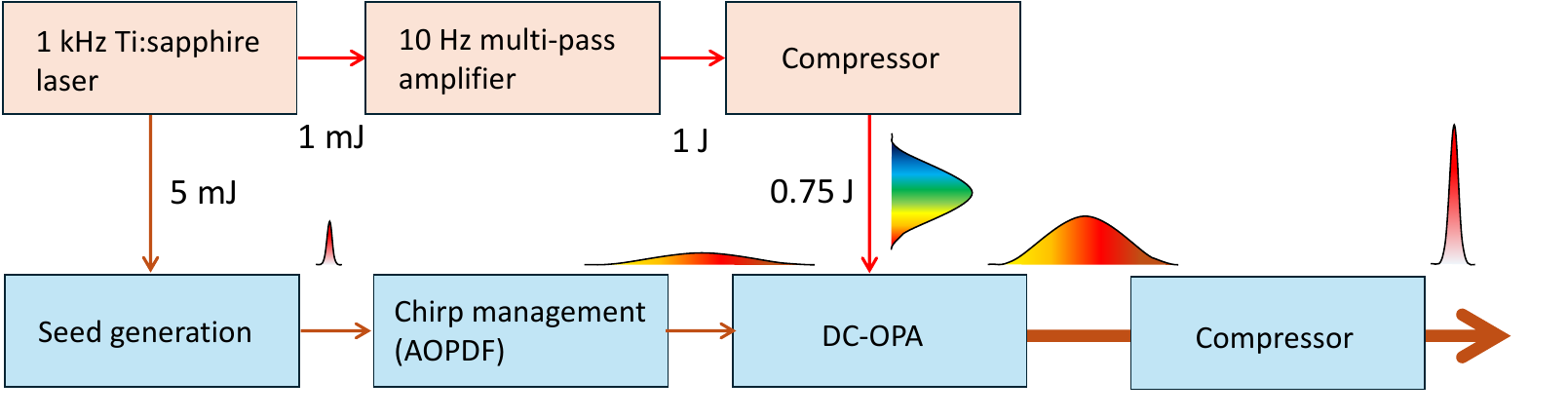}
\caption{Typical configuration of DC-OPA} \label{setup1}
\end{figure}

A conventional configuration of DC-OPA is illustrated in Fig. \ref{setup1}. 
Here, a typical Ti:sapphire CPA system is employed for the front-end laser. 
Part of the output energy before compression (1 mJ) in the CPA system is propagated to the multi-pass amplifier and amplified to 1 J by a four-pass multi-pass amplifier at 10 Hz \cite{pump}. 
The amplified pulse is propagated into the compressor in a vacuum chamber, and the chirp is adjusted by varying the distance between the gratings. 
The remaining 1 kHz part (5 mJ, 30 fs) is used to generate the seed for the DC-OPA. 
The seed is generated via self-phase modulation (SPM) in the sapphire and a two-stage OPA or by SPM in Kr gas via filamentaion \cite{filament} and intra-pulse difference-frequency generation (DFG) \cite{DFG1,DFG2}. 
The generated seed passes through the acousto-optic programmable dispersive filter (AOPDF) with/without Si bulk for chirp management and propagates to the DC-OPA part. Typically, the seed energy is several tens of nJ to several $\mu $J, and is amplified in two- or three-stage DC-OPA to several tens of mJ to approximately 100 mJ. 
The amplified seed is chirped; thus, the chirped seed is compressed by a prism compressor, bulk fused silica or bulk CaF$_2$ considering the wavelength region and chirp amount. 
The phase of compressed pulse is evaluated using the second-order frequency-resolved optical gating (FROG) technique \cite{FROG}. 

Here, we introduce each TW-class DC-OPA system. The configuration and conditions of the DC-OPA are summarized in Table \ref{Configuration_DC_OPA}.
From 1.3 $\mu $m to 1.7 $\mu $m lasers \cite{DCOPA4,1.7umDCOPA} were developed using type I and type II BBO crystals. 
The chirp amount of the 10 $\mu $J seed, generated by a conventional OPA with a Ti:sapphire laser, was adjusted using an AOPDF.
The appropriate pump chirp amount for DC-OPA is positive when amplifying 1.3 $\mu $m--1.5 $\mu $m and negative  when amplifying 1.7 $\mu $m (See Fig. \ref{phase}(a),(b)).
These seeds are amplified by two stages of DC-OPA. 
Dispersion of the amplified seed is compensated by a fused silica prism compressor, and energy greater than 100 mJ is obtained at each wavelength. 
The pulse width was well compressed at each wavelength and reached 44 fs (full width at half maximum: (FWHM)) at 1.5 $\mu $m and 31 fs (FWHM) at 1.7 $\mu $m, nearly at the Fourier transform limit (FTL), resulted in a 2.5 TW and 3.2 TW laser system. 

The 3.3 $\mu $m laser \cite{3umDCOPA} is amplified by type I MgO:LN crystals. 
The seed is generated using DFG after the OPA, and the energy is approximately sub $\mu $J.
The resulting 3.3 $\mu $m seed is positively chirped by the AOPDF and 140 mm Si bulk and amplified by two stages of MgO:LN DC-OPA with  the negatively chirped pump. 
The seed dispersion is compensated by the $\rm{CaF_2}$ bulk, resulting in the pulse width of 80 fs (FWHM), and the pulse energy of 31 mJ, corresponding to an 0.3 TW laser system. 

The PM conditions in Fig. \ref{phase}(d) show that the type I BiBO crystal can amplify the wavelength region where sub-two-cycle laser can be generated \cite{BiBO1,BiBO2,HCO2}. 
To generate the seed for the sub-two-cycle laser, the Ti:sapphire laser is broadened from 0.5$\,\mu $m to 1$\,\mu $m via SPM in Kr gas and this broadened light is used for intra-pulse DFG by the type II BiBO crystal. 
This scheme enables generation of a passively CEP-stabilized MIR seed even if the original laser's CEP is not stabilized. 
The generated seed (10 nJ of energy) passes through the AOPDF and is positively chirped. 
Here, the pump is negatively chirped and the seed is amplified by three stages of the type I BiBO crystals. 
The final output was 105 mJ, pulse width was 10.4 fs (FWHM), and peak power reached 10 TW.

\begin{table}[h]
\caption{Demonstrations of DC-OPA to generate multi-cycle laser (AOPDF: acousto-optic programmable dispersive filter, FS: Fused silica, SPM: self-phase modulation, PM: phase matching, DFG: difference-frequency generation).}
  \centering
  \small 
  \renewcommand{\arraystretch}{1.2} 

  \begin{tabularx}{\textwidth}{lXXXX}
    \hline
    & Ref.\cite{DCOPA4} & Ref. \cite{1.7umDCOPA} & Ref. \cite{3umDCOPA} & Ref. \cite{sub2DCOPA} \\
    \hline
    Seed generation & SPM at sapphire + two-stage OPA & SPM at sapphire + two-stage OPA & SPM at sapphire + two-stage OPA and DFG & Filamentation + intra-pulse DFG \\
    Seed pulse energy ($\mu$J) & 10 & 8 & a few & 0.04 \\
    Chirp management & AOPDF & AOPDF & Si+AOPDF & AOPDF \\
    Seed chirp & positive & positive & positive & positive \\
    Pump chirp & negative & negative & negative & negative \\
    Nonlinear crystal & BBO & BBO & MgO:LN & BiBO \\
    PM & Type II & Type I & Type I & Type I \\
    Amplifier stage & 2 & 2 & 2 & 3 \\
    Pulse energy (mJ)& 110 & 100 & 30 & 100 \\
    Compression & FS prism pair & FS prism pair & CaF$_2$ bulk & FS bulk \\
    Central wavelength ($\mu$m) & 1.5 & 1.7 & 3.3 & 1.7 \\
    pulse width (fs)  & 43 & 31 & 70 & 10.4 \\
    Cycle number & 8.8 & 5.5 & 6.3 & 1.8\\
    CEP (mrad) & N/A & N/A & N/A & 207\\
    Peak power (TW) & 2.5 & 3.2 & 0.3 & 10\\
    \hline
  \end{tabularx}

  \label{Configuration_DC_OPA}
\end{table}

\FloatBarrier
\subsection{Single-shot soft X-ray absorption spectroscopy}
\label{Single-shot soft x-ray absorption spectroscopy using HH beam}
Here, we introduce the high-energy water window HHG \cite{abs_spe_HHG,gascell} using a 1.5 $\mu $m multi-cycle DC-OPA \cite{DCOPA4} and apply near edge X-ray absorption fine structure (NEXAFS) measurement.
Typically, to generate the water window HH using a 1.5 $\mu$m laser system, the driver laser must be tightly focused due to low driver laser energy, and the gas pressure must be increased to several atm to satisfy the PM condition \cite{highpre1,highpre2}. 
With DC-OPA, we can generate an MIR driver laser with a pulse energy that is more than 10 times higher than that of other systems. 
Thus, by taking advantage of the high pulse energy and combining it with the loosely focusing method \cite{loose1,loose2}, we can generate high-efficiency and high-energy water window HH.

\begin{figure}[h]
\centering\includegraphics[width=10cm]{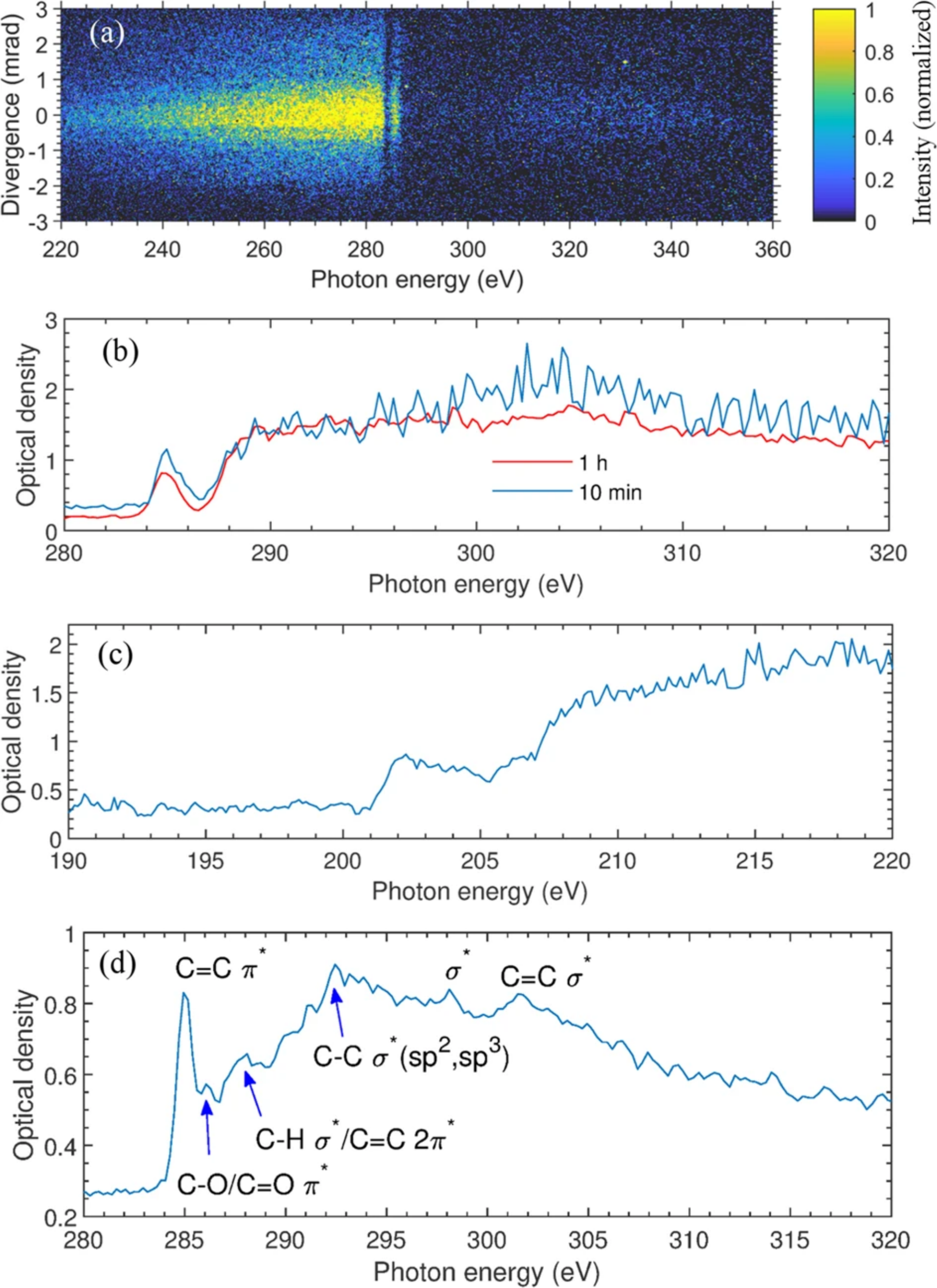}
\caption{NEXAFS by soft X-rays. (a) 2D spectrum after passing through a 1 $\mu $m-thick Mylar film. (b) Absorption spectrum near the carbon K-edge in (a). (c) Absorption spectrum near the chlorine L-edge after passing through a 1.2-$\mu $m-thick Parylene-D film. (d) Absorption spectrum near the carbon K-edge after passing through a 0.25-$\mu $m-thick Parylene-C film. Cited from Ref. \cite{abs_spe_HHG}, \href{https://creativecommons.org/licenses/by/4.0/}{CC BY 4.0.}} \label{absHHG}
\end{figure}


In energy scaling of water window HHG, the DC-OPA pulse energy used for the HHG was 49 mJ with a pulse width of 30 fs, and by focusing the laser using a lens with a focal length of 2 m, the PM condition  was satisfied by suppressing the gas pressure to 1 atm even He. 
Given the loosely focusing method \cite{loose1,loose2} on HHG, the conversion efficiency in the soft X-ray region was improved by more than 10 times compared with previously reported results, and the obtained pulse energy was 3.8 nJ in the water window region, which is more than 100 times higher than a previous study \cite{Con_eff}.
A proof-of-principle experiment of NEXAFS that used a nanojoule coherent water window beam has been demonstrated \cite{abs_spe_HHG}. 
Figure \ref{absHHG} (a) shows the 2D spectrum of the HH after passing through a 1 $\mu$m-thick Mylar film.
After 1 hour, we observed a fine absorption structure near the carbon K-edge (Fig. \ref{absHHG}(a),(b)), and near the chlorine L-edge with 1.2 $\mu$m-thick Parylene-D film(Fig. \ref{absHHG}(c)). 
With a 0.25 $\mu$m-thick Parylene-C film, finer structures in the carbon K-edge were observed in just 2 minutes (Fig. \ref{absHHG}(d)).
Here, only 2$\%$ of the HH pulse energy was utilized to obtain the absorption spectrum; the remaining 98$\%$ was lost in the slit of the spectrometer. 
Thus, we believe  the measurement time can be reduced by approximately 100 times by installing a toroidal mirror to focus the HH beam with the whole energy of the water window regions. 
This novel ultrafast soft X-ray HH driven by DC-OPA opens the door for demonstrating single-shot XAFS and live-cell imaging with a femtosecond time resolution.
Moreover, our high-energy soft X-ray HH will be very useful in applications of XFEL seeding,
nanolithography, ultrafast dynamics studies, and nonlinear soft X-ray physics.

\subsection{Sub-GW IAP generation using a 100-mJ sub-two-cycle DC-OPA}

\begin{figure}[h]
\centering\includegraphics[width=0.8\textwidth]{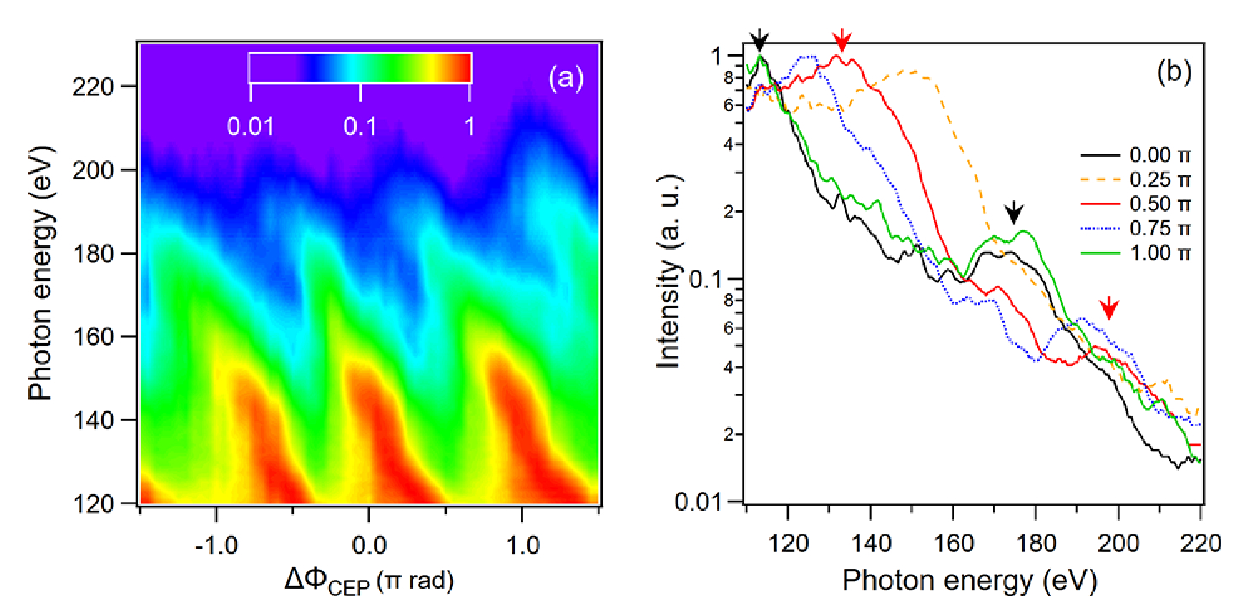}
\caption{(a) CEP dependence of HH spectrum with sub-two-cycle laser. (b) Line profile at each CEP value. Reprinted with permission from \cite{sub2DCOPA} \copyright Optica Publishing Group. } \label{sub2}
\end{figure}

As discussed in the introduction part, the continuum region of the HH spectrum appears at the cutoff region when HH is generated by a few-cycle pulse. 
Multi-TW sub-two-cycle DC-OPA \cite{sub2DCOPA} is not only capable of generating an IAP but also producing a few tens of nanojoule HH due to its high pulse energy.
We test the ability of this multi-TW sub-two-cycle DC-OPA system to generate a sub-GW IAP under downscaled experimental HHG conditions. 
In this evaluation, 5 mJ of energy was focused into a gas cell filled with Ar gas in vacuum, and the CEP dependence of the HH was observed by varying the CEP in steps of 0.05$\pi$ rad. 
Fig. \ref{sub2}(a) shows a 2D map of the CEP dependence of the HH, and Fig. \ref{sub2}(b) shows the HH line profile for each CEP value. 
The fact that the HH spectrum modulates around the cutoff energy suggests that the IAP is generated.
In the line profile (Fig. \ref{sub2}(b)), the half-cycle cutoff (HCO), which typically appears when a sub-two-cycle laser is used as the driver laser \cite{HCO1,HCO2,HCO3},  was also observed.

Given these proof-of-principle experiments, we design an experimental condition that effectively exploits the whole laser energy and show the application prospects of sub-GW IAP generation. 
Since the driver laser energy is very high at 100 mJ \cite{sub2DCOPA}, the beam is expanded to 60 mm in diameter and then focused using a lens with a focal length of 7.6 m
Under this condition, the gas pressure that can satisfy the PM for Ne and He is 0.05 atm and 0.4 atm, the focused intensities are $4.0\times10^{14}\,\rm{W/cm^2} $ and $5.6\times10^{14}\,\rm{W/cm^2} $, and the expected cutoff photon energies are 370 eV and 510 eV, respectively \cite{PM}. 
Based on the driver laser wavelength scaling law of the conversion efficiency of HH \cite{CO_rule3} and the conversion efficiency of previously reported experimental results \cite{gascell}, the pulse energy in the water window region will be  15 nJ for Ne and 5 nJ for He. Assuming a pulse width of approximately 100 as, the peak power of IAP is expected to be a sub-GW level.
Thanks to the nanojoule class soft X-ray HHG by the multi-TW sub-two-cycle DC-OPA, both the single-shot and attosecond temporal resolution can be provided to various ultrafast soft X-ray applications.
\FloatBarrier

\section{TW-class MIR single-cycle laser-based on advanced DC-OPA}
\label{TW-class MIR single-cycle laser based on advanced DC-OPA}

\subsection{Advanced-DC-OPA with heterogeneous nonlinear crystals}

Although conventional DC-CPA has excellent characteristics, e.g., wavelength flexibility and pulse energy scalability, reducing the number of laser cycles in the pulse envelope to less than sub-two cycles is difficult because no nonlinear crystal satisfying the PM condition for DC-OPA over one octave with the pump wavelength of a Ti:sapphire laser is available. 
To overcome this limitation, we developed an advanced DC-OPA method that utilizes two kinds of nonlinear crystals (Fig. \ref{Configuration_DC_OPA}(d)) \cite{xu_single}. 
In this method, nonlinear crystals suitable for different amplified wavelength region are arranged on the collinear axis, and each wavelength region is amplified individually, which makes it possible to amplify single-cycle lasers. 

\begin{figure}[h]
\centering\includegraphics[width=0.6\textwidth]{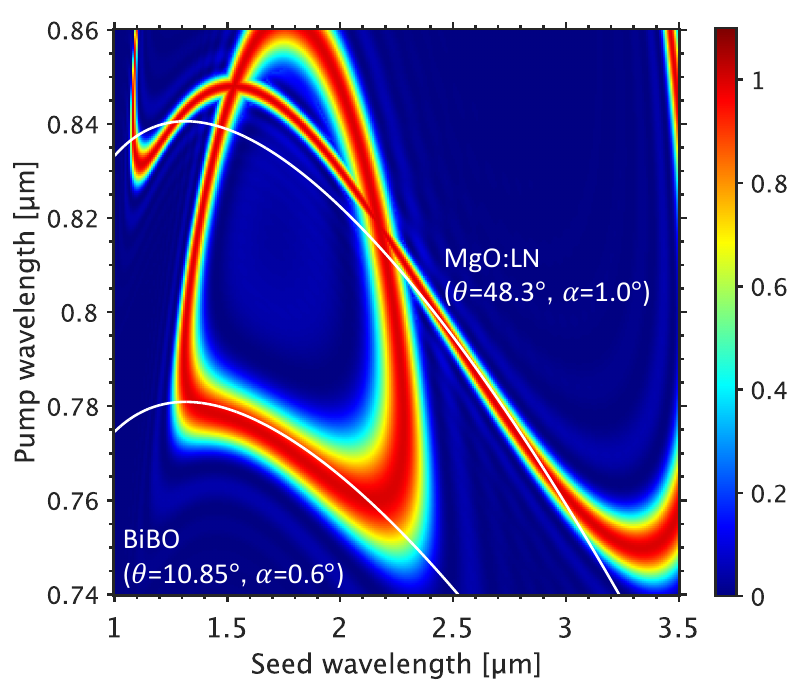}
\caption{PM and chirp matching condition for amplifying a single-cycle laser with type I BiBO ($\theta=10.85^{\circ}$, $\alpha=0.6^{\circ}$), and type I MgO:LN ($\theta=48.75^{\circ}$, $\alpha=1.0^{\circ}$)} \label{PM_single}
\end{figure}

Both the PM and the chirp matching condition should be satisfied to obtain a broadband amplification in the DC-OPA. (See Sec. \ref{Features of DC-OPA}). 
The PM and chirp matching conditions for a single-cycle laser are shown in Fig. \ref{PM_single}. 
We found that the PM condition can be satisfied with a type I BiBO ($\theta =10.85^{\circ}$,$\alpha=0.6^{\circ}$) on the short wavelength region (1.4--2.4 $\mu $m) and type I MgO:LN crystal ($\theta =48.3^{\circ}$, $\alpha=1.0^{\circ}$) on the long wavelength region (2.4--3.0 $\mu $m), where $\alpha$ is the noncollinear angle between the pump and the seed. 
For chirp matching, nonlinear chirp must be considered because the amplification bandwidth is very broad. 
The calculation of the chirp matching is performed assuming that the pump chirp is linear and the seed chirp is the opposite chirp of the material dispersion of the sapphire used for compression. 

The white line in Fig. \ref{PM_single} shows the chirp matching between the pump and the seed. 
When a 100 mm sapphire bulk was used for the pulse compressor, we confirmed that the chirp matching could be satisfied by giving the seed dispersion of $4.5\times10^4\,\rm{fs^2}$. 
As shown in Fig. \ref{PM_single}, it appears that the region where PM can be satisfied with the MgO:LN crystal extends to the short wavelength side (1.4--2.4 $\mu $m) and amplification might be possible with MgO:LN alone; however, chirp matching cannot be satisfied effectively. Thus, amplification was performed on the short wavelength side with the BiBO crystal.

\FloatBarrier
\subsection{Experimental setup of TW-class single-cycle DC-OPA}

\begin{figure}[bh]
\centering\includegraphics[width=0.8\textwidth]{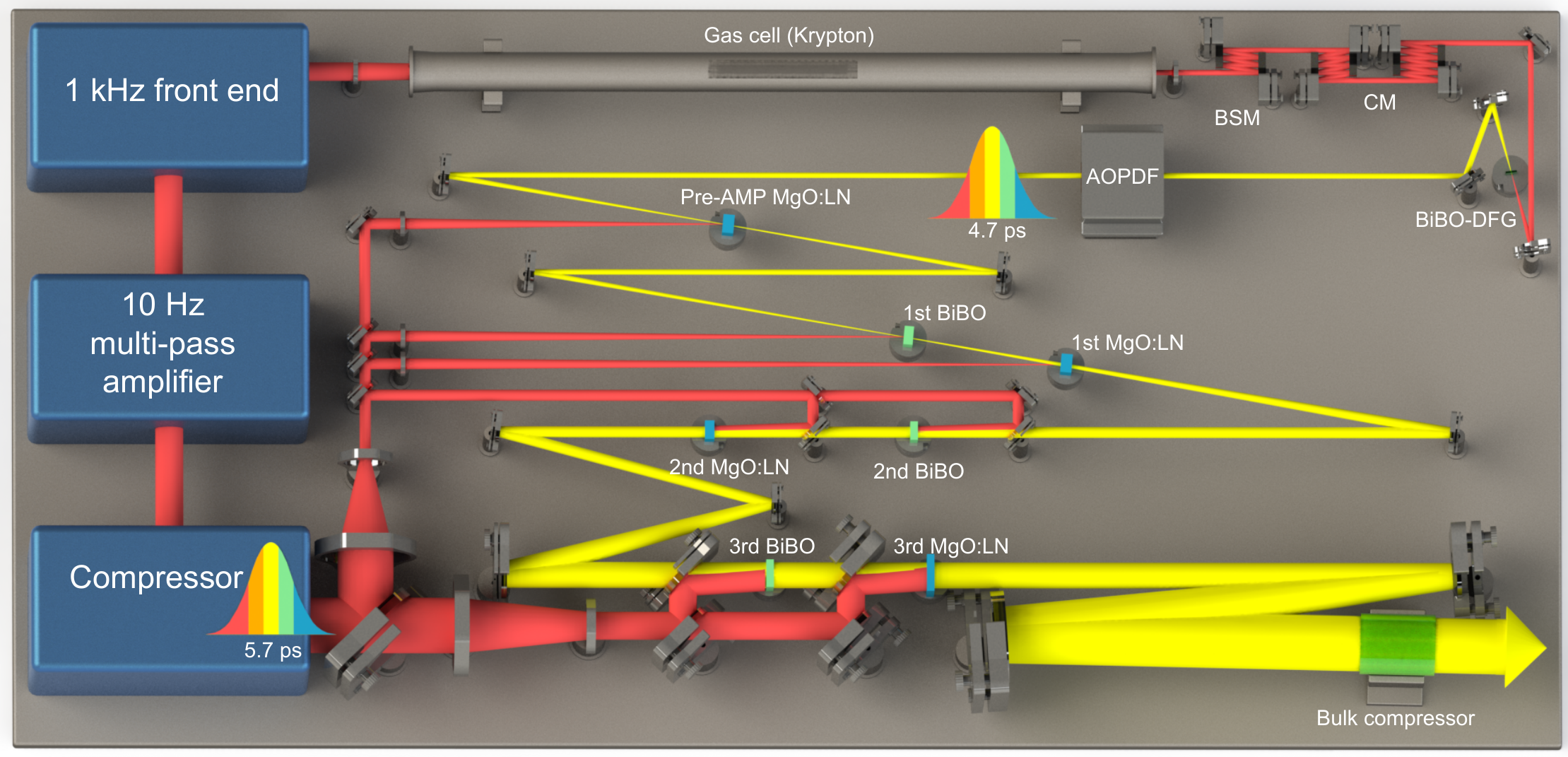}
\caption{Experimental setup for single-cycle DC-OPA (BSM: band-stop mirror, CM: chirped mirror).} \label{setup_single}
\end{figure}

The top view of the experimental setup for the single-cycle DC-OPA system is shown in Fig. \ref{setup_single}. 
Here, the front-end 1 kHz Ti:sapphire laser and the 10 Hz Ti:sapphire multi-pass amplifier used as the pump are the same as those in the multi-cycle DC-OPA \cite{DCOPA3} setup. 
Here, to generate the seed, the Ti:sapphire laser bandwidth was broadened from 0.5$\,\mu $m to 1.0$\,\mu $m via SPM in Kr gas. Then, after reflection through chirped mirrors and a band-stop mirror \cite{HCO2}, a CEP-stabilized seed was generated via intra-pulse DFG in a type II BiBO crystal $(\theta=57^{\circ})$. 
The generated seed was passed through an AOPDF and amplified using the advanced DC-OPA configuration.

\FloatBarrier

The DC-OPA comprises seven amplifier stages: MgO:LN pre-amplifier, three-stage MgO:LN amplifier, and three-stage BiBO amplifier. 
The MgO:LN pre-amplifier compensates for the attenuation due to absorption above 2.7 $\mu $m in the BiBO crystal and the low quantum efficiency when amplifying the long wavelength side. 
In the pre-amplifier and the first-stage amplifier, the pump and the seed are focused to nonlinear crystals because the energy is less than 1 mJ. 
In the second- and third-stages, the energy is greater than 10 mJ; thus, the pump is down-collimated. 
However, Galileo-type down-collimation is employed to avoid ionization of the air with Kepler type down collimator, and, if the optical path after down-collimation is too long, the beam quality deteriorates considerably; thus, the optical path was designed to be as short as possible after down-collimation. 
A 40-mm sapphire bulk was used as a pulse compressor, and the inverse dispersion of this sapphire, three BiBOs, four MgO:LNs and seven $\rm{CaF_2}$s for the dichroic mirror in each amplifier was given by AOPDF. 
Here, the seed pulse width in the amplifier was approximately 4.7 ps, and the pump pulse width is adjusted to 5.7 ps by changing the distance between the two gratings in the vacuum compressor. 
Note that the pump and the seed pulse width differed because not all the wavelength region of the pump can be used for DC-OPA. 
The thickness in the pre-amplifier and the three subsequent amplifier stages are 6 mm, 6 mm, 5 mm, and 4 mm for MgO:LN, and 5 mm, 4 mm, and 4 mm for BiBO. 
BiBO, which is used for the amplification of the short wavelength side, has high quantum efficiency and a high nonlinear optical constant; thus, its thickness is less than that of MgO:LN.

\subsection{Experimental results}

\begin{figure}[h]
\centering\includegraphics[width=0.8\textwidth]{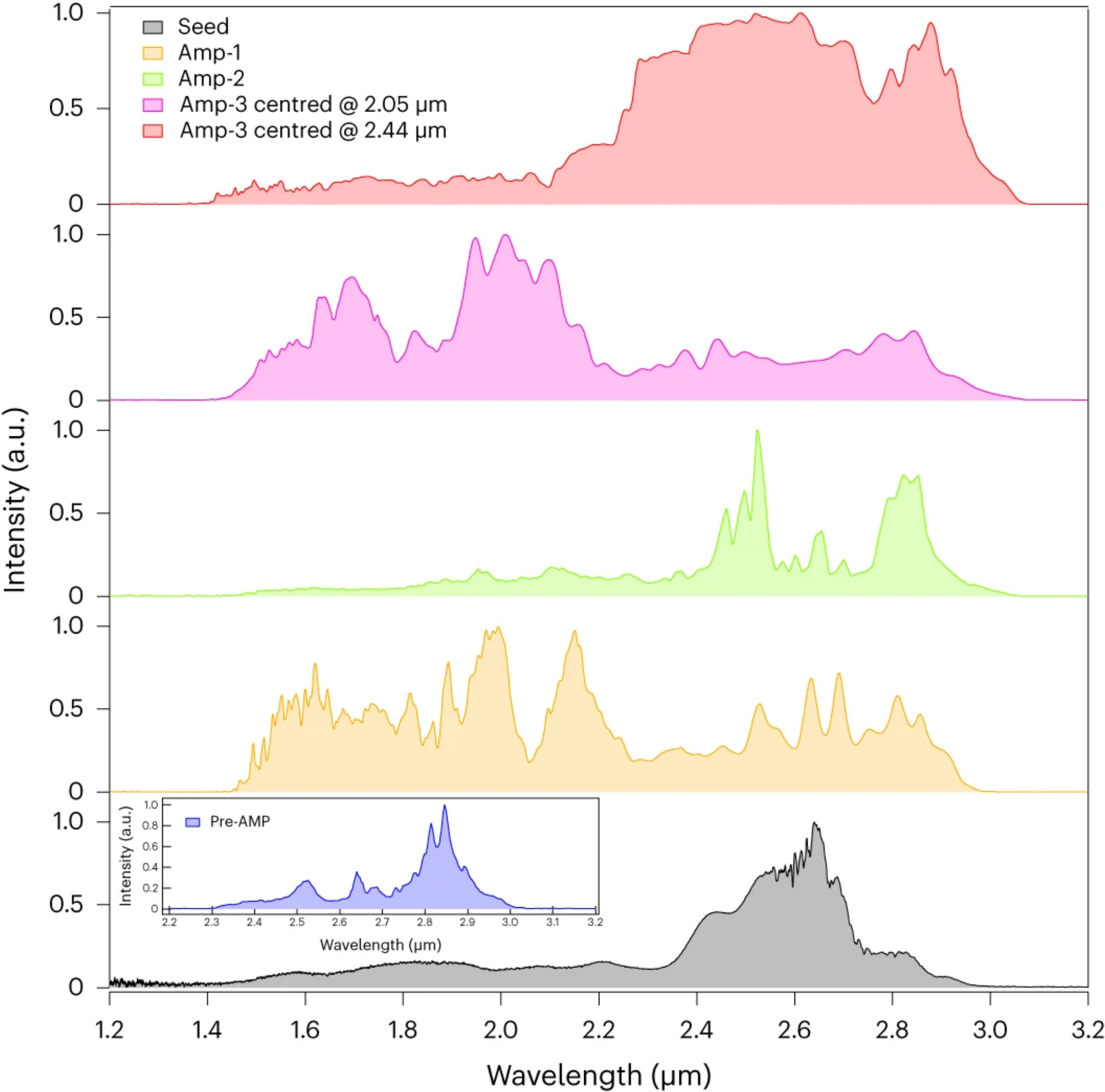}
\caption{Spectrum at each stage amplifier. From the bottom: the seed (black), the first-stage (orange), the second-stage (green), the third-stage centered at 2.05 $\mu $m(pink), and the third-stage centered at 2.44 $\mu $m(red). The inset (blue) shows the amplified spectrum after the pre-amplifier. Cited from Ref. \cite{xu_single}, \href{https://creativecommons.org/licenses/by/4.0/}{CC BY 4.0.}} \label{singleamp_spe}
\end{figure}

Figure \ref{singleamp_spe} shows the amplification spectrum in each stage amplifier, where the bottom black line is the seed spectrum. 
A seed from 1.4 $\mu $m to 3.0 $\mu $m was generated, and the inset (blue line) shows the spectrum amplified by the MgO:LN pre-amplifier. 
In the first-stage amplifier, 45 $\rm{GW/cm^2}$ and 40 $\rm{GW/cm^2}$ are is incident on BiBO and MgO:LN respectivly. 
The long wavelength side is amplified by the pre-amplifier; thus, the spectrum is uniform even though the pump intensity to MgO:LN is low. 
In the second-stage amplifier, 40 $\rm{GW/cm^2}$ and 50 $\rm{GW/cm^2}$ are incident on BiBO and MgO:LN respectivly. 
Thus, the long wavelength side is relatively strong. 
In the third-stage amplifier, the central wavelength can be varied by changing the ratio of the pump intensity to the BiBO and MgO:LN. 
When the pump intensity was 50 $\rm{GW/cm^2}$ to BiBO and 40 $\rm{GW/cm^2}$ to MgO:LN, the central wavelength was 2.05 $\mu $m, and the energy was 61 mJ. When the pump intensity ratio was reversed, the central wavelength was 2.44 $\mu $m, and the energy was 53 mJ. 
The output energy decreases when the pump intensity to BiBO is reduced because BiBO has a higher nonlinear coefficient compared with MgO:LN. 

\begin{figure}
\centering\includegraphics[width=0.8\textwidth]{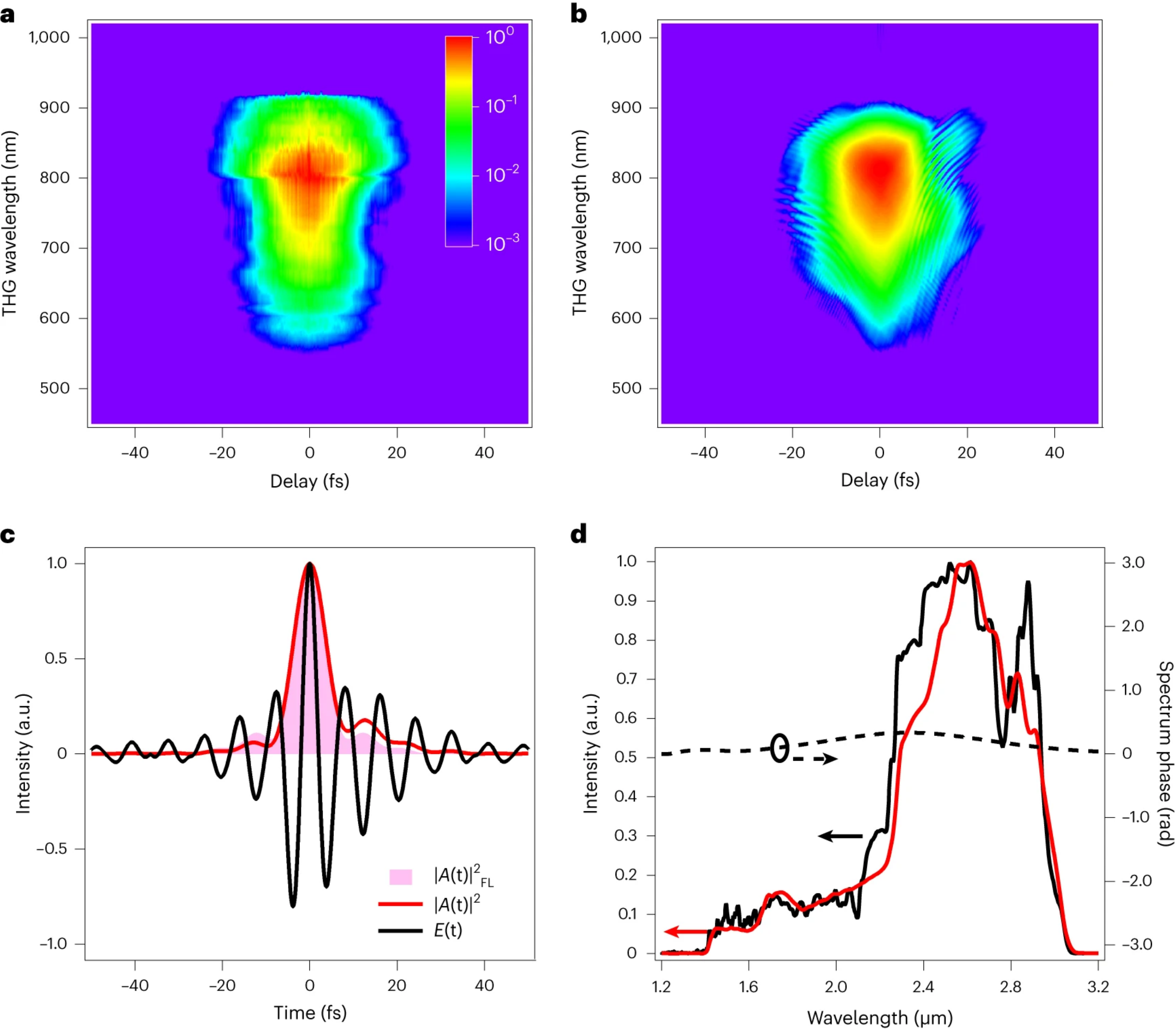}
\caption{THG-FROG results of the single-cycle laser. (a) Measured and (b) reconstructed THG-FROG traces. (c) Reconstructed temporal profile and (d) reconstructed spectrum (red solid line) and spectral phase (black dashed line). The red line is the measured spectrum. Cited from Ref. \cite{xu_single}, \href{https://creativecommons.org/licenses/by/4.0/}{CC BY 4.0.}
} \label{THGFROG}
\end{figure}

\begin{figure}
\centering\includegraphics[width=0.6\textwidth]{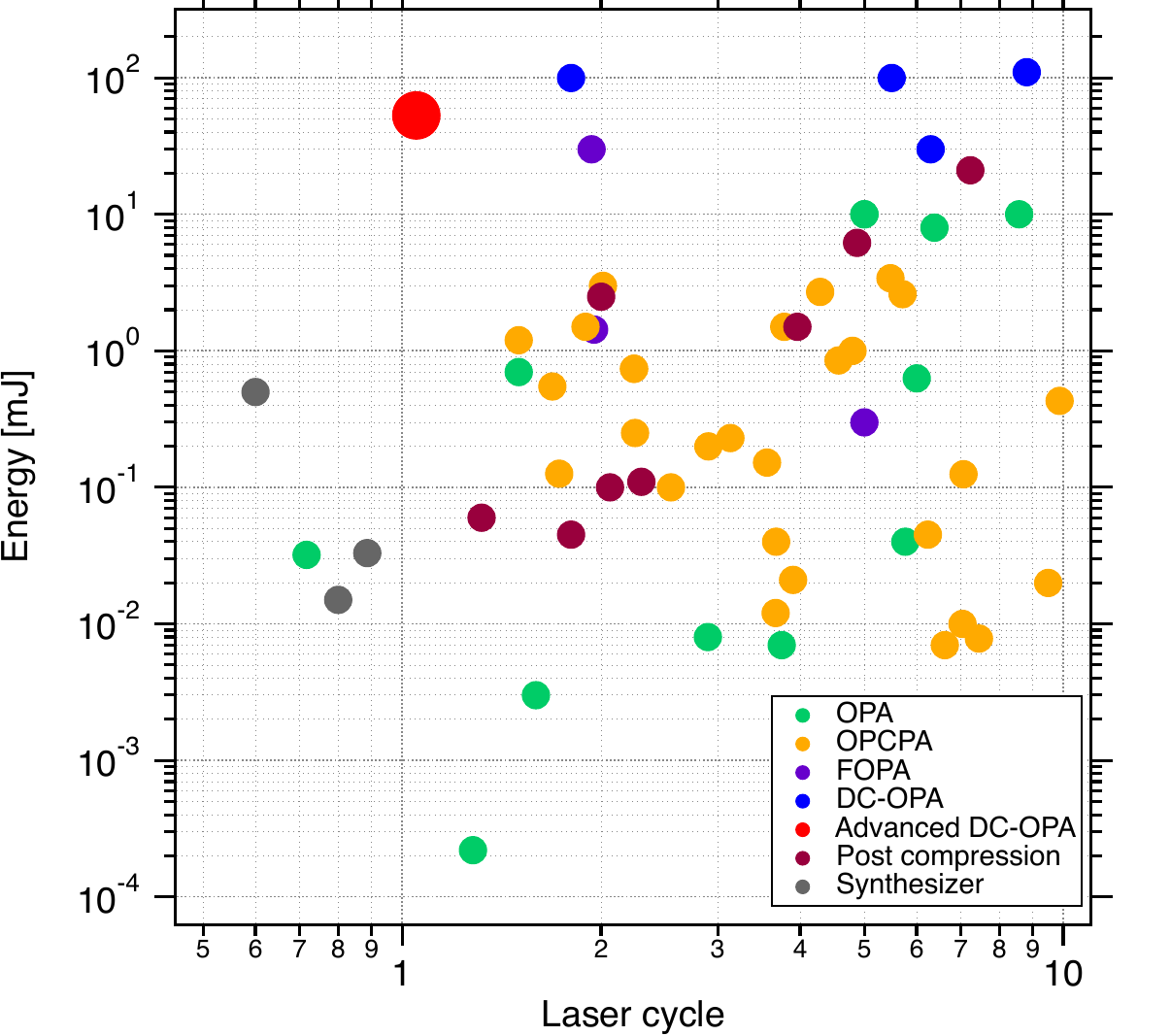}
\caption{Development status of MIR laser systems with the number of laser cycles and laser pulse energy. \cite{OPAref1,OPAref4,OPAref5,OPAref6,OPAref7,OPAref8,OPAref9,OPAref10,OPAref11,OPAref12,OPAref13,OPAref58}, OPCPA \cite{OPCPAref21,OPCPAref22,OPCPAref23,OPCPAref24,OPCPAref26,OPCPAref28,OPCPAref29,HCO2,OPCPAref23,OPCPAref32,OPCPAref33,OPCPAref34,OPCPAref35,OPCPAref36,OPCPAref37,OPCPAref38,OPCPAref39,OPCPAref40,OPCPAref42,OPCPAref44,OPCPAref45,OPCPAref46,OPCPAref47,OPCPAref49,OPCPAref51,OPCPAref52}, FOPA \cite{FOPA1,FOPA2}, DC-OPA \cite{DCOPA4,3umDCOPA,1.7umDCOPA,sub2DCOPA}, advanced DC-OPA \cite{xu_single}, post-compression \cite{PostCompref61,PostCompref63,PostCompref64,PostCompref65,PostCompref66,PostCompref67,PostCompref69,elu2017high}, and a waveform synthesizer \cite{syn3,synref16,synref17}. 
} \label{lasersystem}
\end{figure}

After amplification, the amplified beam is expanded by the convex and concave mirrors, and dispersion is compensated by passing through the sapphire bulk compressor. The pulse width was evaluated by the third-order harmonic generaion (THG)-FROG method using surface THG from a 200 nm-thick $\rm{Si_3N_4}$ \cite{THGFROG}. 
The measured FROG trace is shown in Fig. \ref{THGFROG}(a).
The FROG error of the reconstructed FROG trace (Fig. \ref{THGFROG}(b)) is sufficiently small at 0.78$\%$, and the reconstructed spectrum matches well with the spectrum directly measured using a scanning-type spectrometer (Fig. \ref{THGFROG}(d)).
The phase is nearly flat in the wavelength region from 1.4$\,\mu $m to 3.0$\,\mu $m, the pulse width is 8.58 fs (FWHM), and the central wavelength is 2.44$\,\mu $m; thus, the number of laser cycles is 1.05, resulting in a 6 TW single-cycle laser system.
We also evaluated beam quality and CEP stability, which are important in HHG experiment. 
Beam quality was evaluated by focusing the beam using a concave mirror (0.5 m focal length).
The $\rm{M^2}$ value was 1.24 and 1.29 in the horizontal and verticaßl directions, respectively. 
The CEP stability was measured using an $f$-$2f$ interferometer \cite{f2fsingle}. 
Due to the one octave spectrum bandwidth, the CEP can be evaluated by generating the second harmonic at $\rm{LiIO_3}$, aligning the polarization with a polarizer, and measuring the spectrum. 
Here, the single-shot stability of the CEP value was 228 mrad (root mean square).
As a result, the advanced DC-OPA method enabled amplification of a one octave spectrum bandwidth, and we succeeded in developing a laser system with the highest energy of any single-cycle laser (See Fig. \ref{lasersystem}).

\FloatBarrier

\FloatBarrier

\subsection{Supercontinuum HHG by single-cycle DC-OPA}

Here, we introduce one octave spanning super continuum HHG as an application of TW-class single-cycle laser \cite{singleHHG}. 
As explained at the introduction, there is a correlation between the number of driver laser cycles and the percentage of the continuum region in the HH spectrum, and the fewer laser cycles, the broader the continuum region appears \cite{xu_single}.
Using a single-cycle driver laser, it is possible to make 40$\%$ continuum region for the entire HH spectrum, which is advantageous in terms of generating shorter attosecond pulses. 
To demonstrate this, we performed HHG using a single-cycle driver laser. 

\begin{figure}
\centering\includegraphics[width=0.9\textwidth]{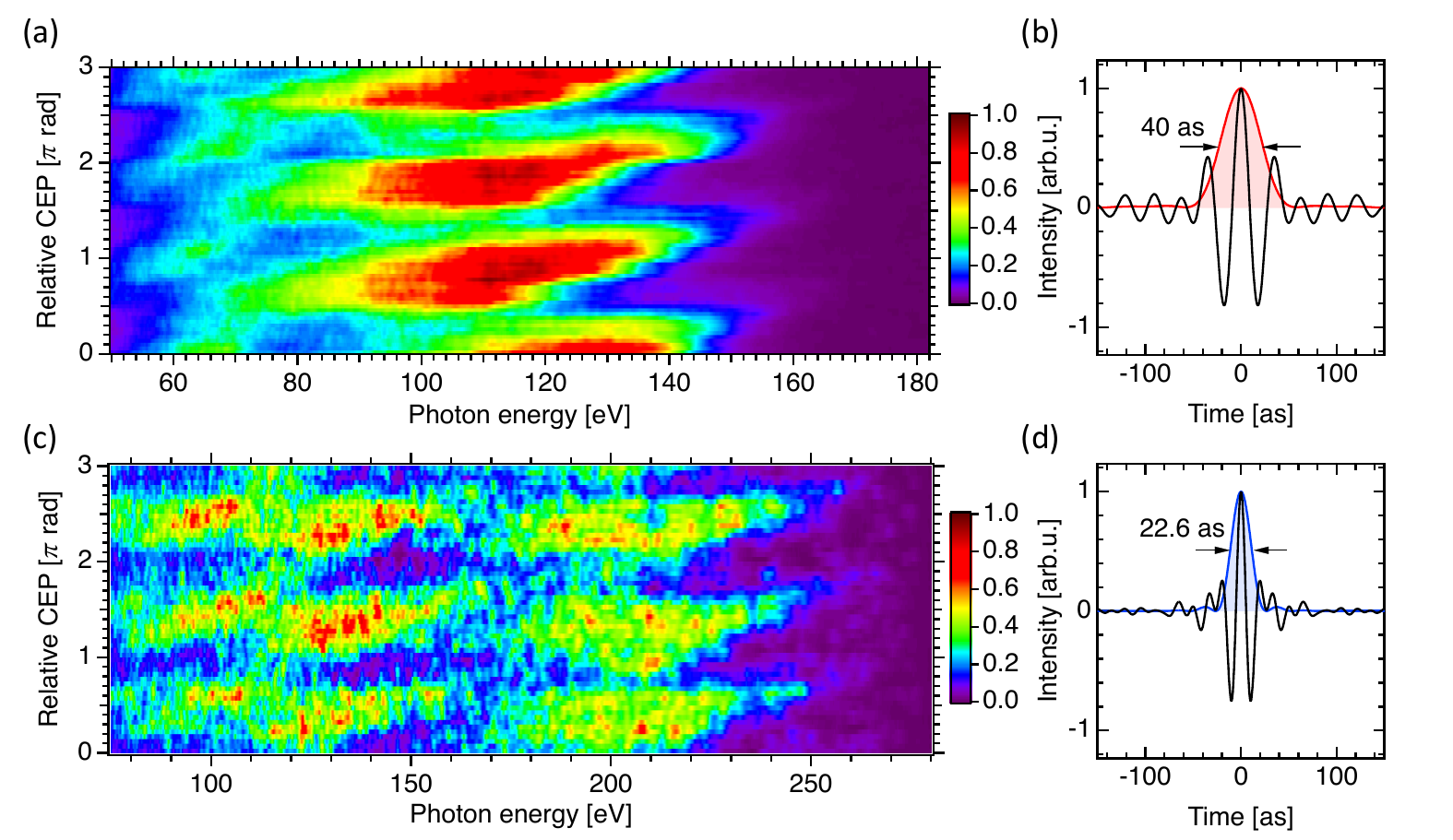}
\caption{CEP dependence of experimental HH spectrum by single-cycle laser with (a) Ar gas and (c) Ne gas, and their respective FTL pulse widths (b) and (d). Adapted with permission from \cite{singleHHG} \copyright Optica Publishing Group.
} \label{continuumHHG}
\end{figure}

The single-cycle laser was focused to the gas cell (10 mm in length) using a concave mirror (1.5 m focal length). 
In this experiment, Ar gas and Ne gas were used as the interaction media with the pulse gas cell \cite{gascell}. 
The focusing intensity and backing pressure was set to $1.3\times10^{14}\,\rm{W/cm^2} $ and 0.6 atm for the Ar gas and $2.5\times10^{14}\,\rm{W/cm^2} $ and 6.0 atm for the Ne gas.
The CEP dependence of the HH spectrum is illustrated in Fig. \ref{continuumHHG}(a),(c). 
When a single-cycle driver laser was used for both the Ar and Ne gases, a continuum region of approximately 40 $\%$ of the entire HH spectrum was observed, as evaluated in Fig. \ref{cycyle vs cont}. 
With the Ne gas, the HCO frequency and secondary diffraction signals were also observed. 
The HH spectrum has a one-octave continuum region (like the driver laser) which is not only an IAP but also a single-cycle soft-x-ray attosecond pulse \cite{singleHHG}.
To create the shorter pulse duration of the IAP, the atto-chirp on the HH should be considered.
Calculations were performed to determine whether the atto-chirp that typically accompanies HH generated from gases can be compensated \cite{attochirp1,attochirp2}. 
We found that the atto-chirp can be compensated using a 207 nm Zr filter for HH from Ar and a 278 nm Sn filter for HH from Ne, and a single-cycle attosecond pulse with 1.1 cycle, 40 as at 118 eV (Fig. \ref{continuumHHG}(b)) and 1.1 cycle, 23 as at 206 eV (Fig. \ref{continuumHHG}(d)) can be generated.
IAP generation using a single-cycle driver laser is expected to facilitate a sub-10 as pulse or even zepto-second pulse generation.
In addition, this soft X-ray single-cycle attosecond pulse is expected to have high utility for ultrafast science \cite{mumkick1,mumkick2} and various applications in quantum information science \cite{QIP}.

\section{Future prospect}
\label{Future prospect}

Figure \ref{lasersystem} shows the development status of MIR laser based on several methods. 
The advanced DC-OPA has produced the most energetic among single-cycle laser systems; however, DC-OPA has not yet reached the ability limit on the number of laser cycles.
In addition, the average power of DC-OPA can be improved using a proper pump laser rather than a Ti:sapphire laser.
Here, we introduce the prospect and the proof-of-principle experiment of developing a TW-class sub-cycle laser \cite{subnishi}  by expanding the amplification bandwidth from one octave of single-cycle lasers to 1.5 octaves (Sec.  \ref{TW-class sub-cycle DC-OPA pumped by Ti:sapphire laser}).
To increase the average power of DC-OPA , fiber-based lasers and/or  thin-disk lasers (TDL) can be employed for DC-OPA.
In particular, with recent developments in TDL technology, a few hundred watts of the average power with a few hundred femtosecond pulse width has been commercially available.
By combining TDL with post-compression technology and using the DC-OPA method, a single-cycle laser with high-average power can be obtained. 
Here, we discuss the prospects of DC-OPA pumped by a Yb-based TDL (1 $\mu$m) and a Ho-based TDL (2 $\mu$m) in Sec. \ref{DC-OPA pumped by TDL}.

\subsection{Sub-cycle DC-OPA pumped by Ti:sapphire laser}
\label{TW-class sub-cycle DC-OPA pumped by Ti:sapphire laser}

\begin{figure}
\centering\includegraphics[width=0.6\textwidth]{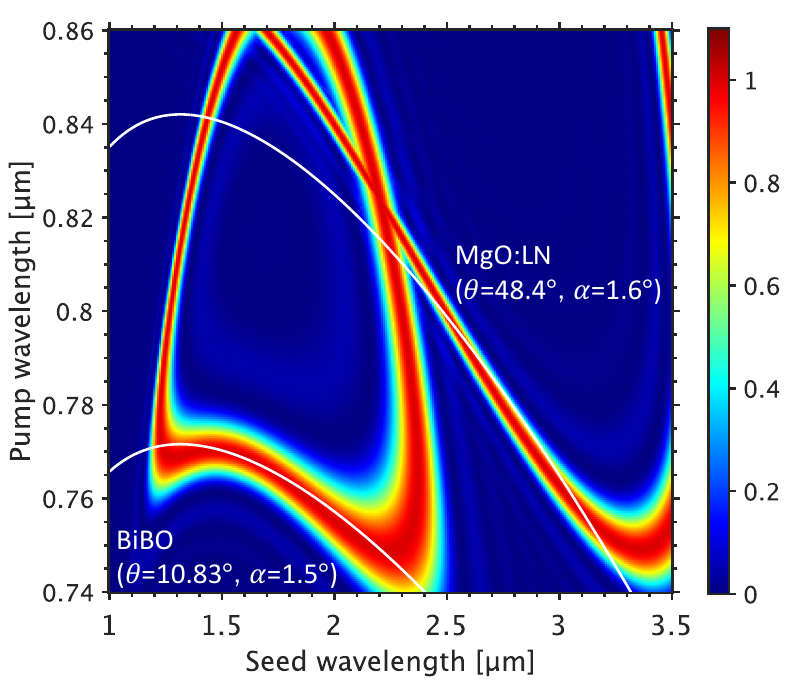}
\caption{PM and chirp matching condition for sub-cycle laser with BiBO crystal ($\theta =11.0^{\circ}$,$\alpha=0.6^{\circ}$) and MgO:LN crystal ($\theta =48.4^{\circ}$,$\alpha=1.6^{\circ}$)
} \label{subcycle PM}
\end{figure}

We introduce a proof-of-principle experiment to demonstrate a TW-class sub-cycle laser that surpasses the TW-class single-cycle lasers. 
To achieve a sub-cycle laser, $\sim$1.5 octave amplification bandwidth is required; thus, we reconsider the PM and chirp matching condition. 
The PM and chirp matching condition for the 1.5 octave DC-OPA is shown in Fig. \ref{subcycle PM}. 
For wavelength region of 1.2 $\mu $m to 2.5 $\mu $m, a BiBO crystal ($\theta =11.0^{\circ}$,$\alpha=0.6^{\circ}$) is used, and for the wavelength region of 2.5 $\mu $m to 3.2 $\mu $m, an MgO:LN crystal ($\theta =48.4^{\circ}$,$\alpha=1.6^{\circ}$) is used.
In the case of DC-OPA pulse compression with a 100 mm sapphire bulk, the chirp matching shown by the white line in Fig. \ref{subcycle PM} can be satisfied by setting the pump dispersion to $4.8\times10^{4}~\rm{fs^2} $.

\begin{figure}
\centering\includegraphics[width=0.9\textwidth]{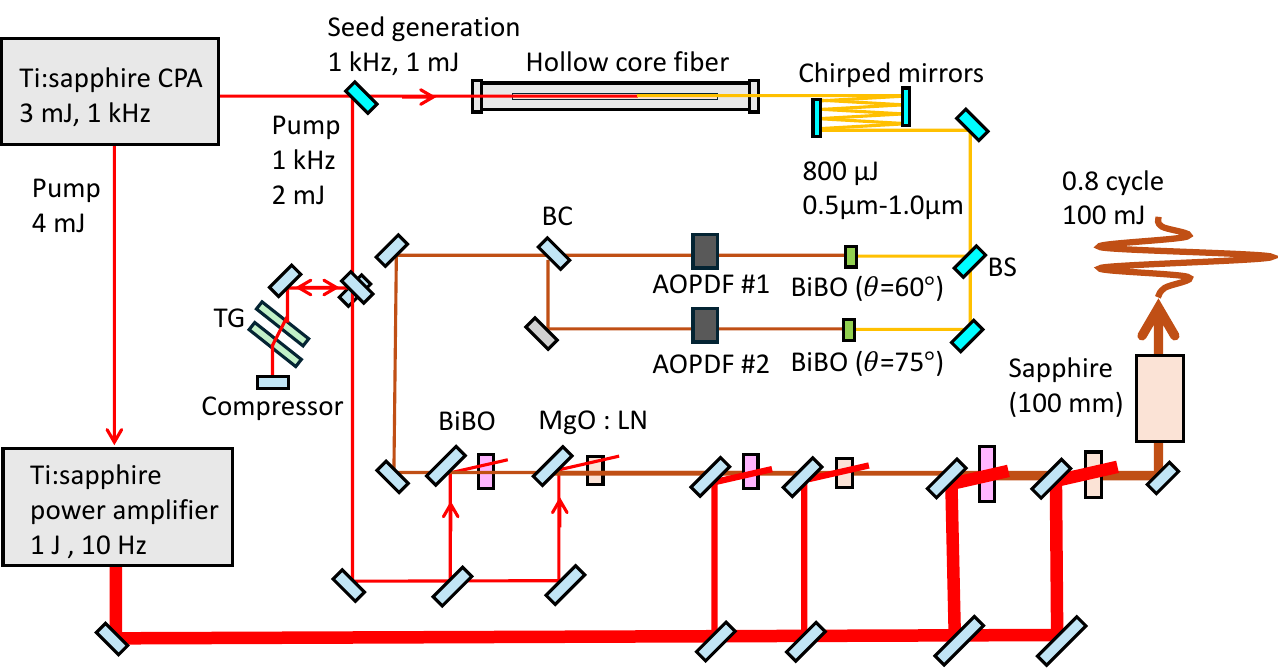}
\caption{Conceptual setup of TW-class sub-cycle laser (BS: beam splitter, BC: beam combiner, TG: transmission-type grating).
} \label{subcycle setup}
\end{figure}

Figure \ref{subcycle setup} shows the conceptual setup for a TW-class sub-cycle DC-OPA, where several different techniques from single-cycle DC-OPA laser system are used. 
Here, the front-end is a Ti:sapphire laser, which is the same as the single-cycle DC-OPA, is used to generate the pump and seed \cite{pump}. 
For seed generation, the spectrum from front-end Ti:sapphire laser system is broadened via hollow core fiber (HCF). Conventionally, filamentation in Kr gas is employed for SPM; however, with HCF, the spectrum is broadened uniformly. 
Thus, the energy available for intra-pulse DFG increases by several tens of times.
After HCF, the dispersion of the light for the seed generation is compensated by chirped mirrors, and then the CEP-stabilized seed is generated via intra-pulse DFG. 
In this configuration, a BiBO crystal with two different optical axes is employed, and a broadband (0.5--1.0 $\mu $m) beam splitter is inserted before the DFG. Then, DFG is performed in two crystals.
If one BiBO crystal is employed to generate the 1.5 octave seed, the crystal must be thinned to approximately several tens of $\mu $m. As a result, the obtained seed energy will be low. 
Thus, two BiBO crystals with different cutting angle are used, taking advantage of the high energy of several hundred $\mu $J obtained by HCF. 
Here, a BiBO type II crystal $(\theta =60^{\circ})$ was adjusted to generate the wavelength region of 1.2 $\mu $m to 2.5 $\mu $m, and another type II BiBO crystal $(\theta =75^{\circ})$ was adjusted to generate the wavelength region of 2.5 $\mu $m to 3.2 $\mu $m. 
The obtained seed was then passed through an AOPDF to control its dispersion; however, the AOPDF bandwidth is limited to approximately one octave due to its specification. 
Thus, two kinds of AOPDF must be employed \cite{subopa}. 
The two dispersion-controlled seeds were then combined in a specially designed beam combiner and amplified using the advanced DC-OPA method. 

Note that the amplification process involves three-stage amplifier. 
The first-stage amplifier uses a 1 kHz pump ($\sim$2 mJ), and the second- and third-stage amplifiers use 10 Hz pump (100 mJ, 1 J). 
The 2 mJ remaining from the pulse for the seed generation is chirped by another transmission-type grating compressor. 
By employing a 1 kHz laser in the first-stage amplifier, the optical path length of the 10 Hz laser can be reduced, which saves space and prevents beam quality degradation. 
The beam diameter is adjusted in the second and third stages while paying careful attention to the depletion \cite{OPAdeplen} or parametric fluorescence in OPA. 
Assuming $\sim$6$\%$ conversion efficiency, the expected output is a 10 TW sub-cycle laser with 60 mJ pulse energy, 5.8 fs (FWHM) pulse width, 2.15 $\mu $m central wavelength, and 0.81 cycles (Table \ref{Prospect of the DC-OPA}).

\begin{table}[h]
\caption{Prospects of DC-OPA (CE: conversion efficiency, TDL: thin-disk laser).}
\label{Prospect of the DC-OPA}
  \centering
   \small 
   {
    \begin{tabular}{lccc}
        \toprule
        Pump for the DC-OPA & Ti:sapphire (0.8 $\mu$m) & Yb-TDL (1 $\mu$m) & Ho-TDL (2 $\mu$m) \\
        \midrule
        Nonlinear crystal & BiBO+MgO:NL  & LiIO$_3$  & $\rm{LiInSe_2}$ \\
        Wavelength range ($\mu$m) & 1.2--3.2   & 2.0--4.4  & 3.5--7.0  \\
        Bandwidth (octave) & 1.40 & 1.14 & 1.0\\
                FT pulse width (fs) & 5.8 & 11.0 & 20.6\\
        Laser cycle & 0.81 & 1.22 & 1.29 \\
        CEP & Stabilized & Stabilized & Stabilized \\
        Expected CE (\%) from pump &  6.5 $\%$ \cite{xu_single} &  4.3 $\%$ \cite{LiIO_CE} & 2.6 $\%$ \cite{LiInSe2_CE} \\
        \bottomrule
    \end{tabular}%
    }
\end{table}

\begin{figure}[h]
\centering\includegraphics[width=10cm]{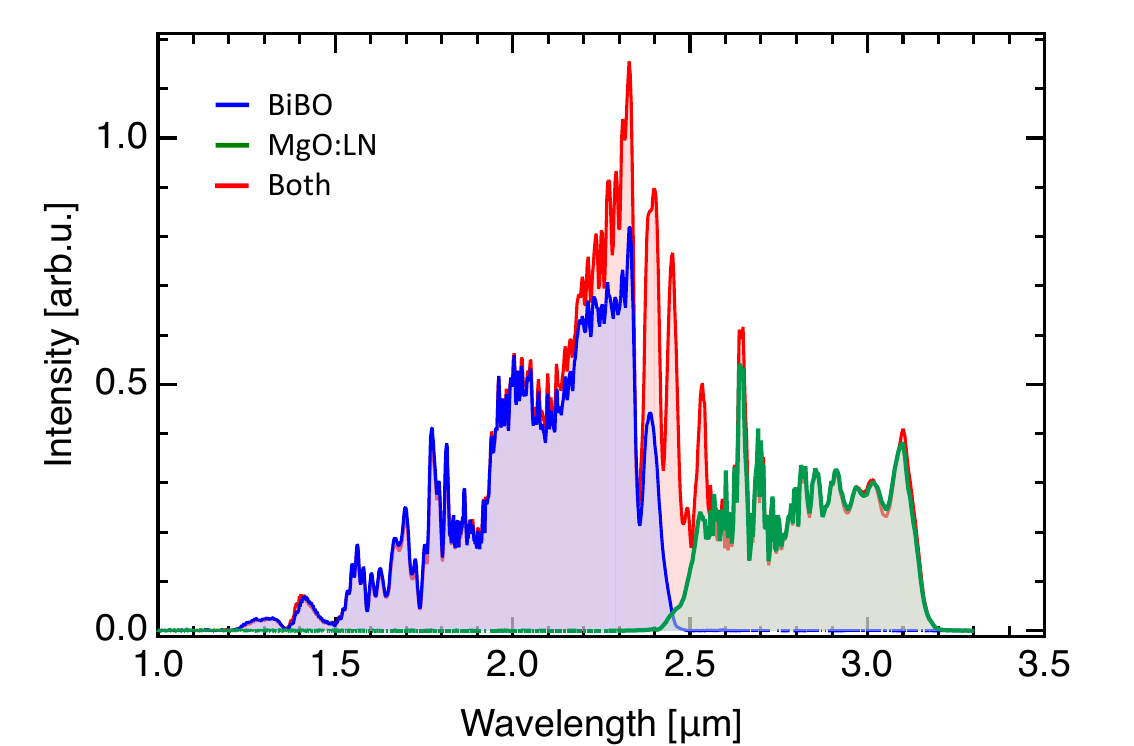}
\caption{Amplified spectrum for the sub-cycle laser by BiBO (blue), MgO:LN (green), and both (red) crystals via DC-OPA.} \label{subcycle amp}
\end{figure}

Based on the conceptual design, we confirmed the amplification bandwidth of DC-OPA in the first-stage amplifier \cite{subnishi}.
In Fig. \ref{subcycle amp}, the blue, green, and red filled lines show the amplified spectrum using the BiBO crystal, the MgO:LN crystal, and both crystals, respectively. 
Here, the two seeds energy after the AOPDF was approximately 100 nJ.
As predicted by our calculations, we found that it is possible to amplify the wavelength region of 1.2$\,\mu $m to 3.2$\,\mu $m with energy of 15$\,\mu $J. 
When both crystals were used simultaneously, an interference signal was observed at approximately 2.5$\,\mu $m; however, this interference signal appears because the delay between the two seed components is not completely adjusted, and this interference signal disappears when the delay is 0.
If the sub-cycle DC-OPA system is applied to attosecond pulse generation via HHG, approximately 70$\%$ of the entire HH spectrum will be in the continuum, and it should be possible to generate an attosecond pulse with an FTL of sub-10 as.

\subsection{High average power DC-OPA pumped by 1 $\mu$m and 2 $\mu$m lasers }
\label{DC-OPA pumped by TDL}

DC-OPA method can be employed universally for the energy scaling of near-IR, MIR, and far-IR pulses regardless of the kind of nonlinear crystal, and it facilitates efficient generation of few-cycle CEP stabilized MIR pulses with TW-class peak power.
Note that the DC-OPA method can be expanded in the case of other pump wavelengths by selecting appropriate nonlinear crystals.
Recently, thanks to the development of Yb- or Ho-based laser and advances in TDL technology, the development of lasers at the hundreds of watts to kW levels is progressing \cite{YbTDL1,YbTDL2,YbTDL3,HoTDL}. 
In the following, we discuss a high average power, single-cycle DC-OPA at the 3$\,\mu $m region using the Yb-based TDL and at the 5$\,\mu $m region using the Ho-based TDL with post-compression. 

\begin{figure}
\centering\includegraphics[width=0.6\textwidth]{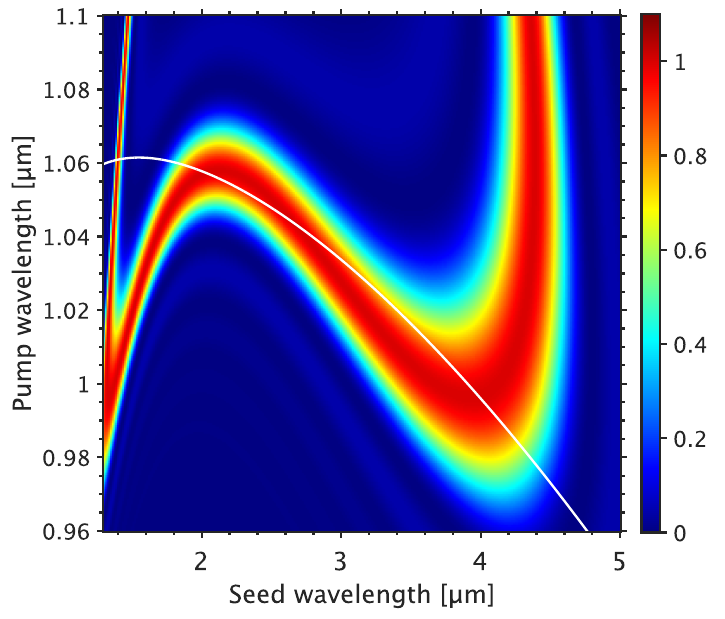}
\caption{PM and chirp matching for single-cycle laser by Yb-based TDL with type II $\rm{LiIO_3}$$(\theta =17.25^{\circ})$.} \label{Yb PM}
\end{figure}

The PM and chirp matching condition of DC-OPA pumped by Yb-based TDL (1$\,\mu $m) for generating a high average power  3$\,\mu $m single-cycle pulse is shown in Fig. \ref{Yb PM}.
Here, the nonlinear crystal is a type I $\rm{LiIO_3}$ $(\theta =17.25^{\circ})$ crystal. 
In Fig. \ref{Yb PM}, the white line shows the chirp matching between the pump and the seed. 
When a $\rm{CaF_2}$ bulk compressor (40 mm thick) was used and the pump dispersion was set to $\rm{4.5 \times 10^4 ~fs^2}$, wavelength region of 2.0$\,\mu $m to 4.4$\,\mu $m could be amplified simultaneously. 
To efficiently cover this PM region of DC-OPA, the spectrum bandwidth of the Yb-TDL must be broadened because its spectrum bandwidth is typically several nm.
To broaden the spectrum bandwidth of Yb-TDL, the multi-pass gas cell technique will be a promising method \cite{highpowerPC1,highpowerPC2}.

\FloatBarrier

\begin{figure}
\centering\includegraphics[width=0.8\textwidth]{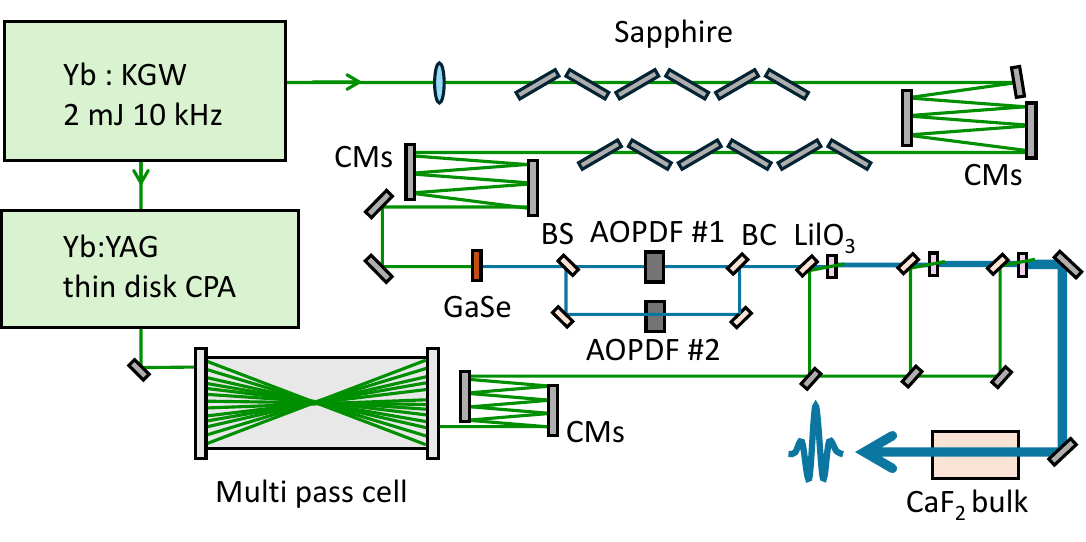}
\caption{Conceptual experimental setup to demonstrate a single-cycle pulse at 2.7 $\mu$m pumped by the Yb-based TDL DC-OPA (BS: beam splitter, BC: beam combiner, CMs: chirped mirrors)}
\label{Yb setup}
\end{figure}

The conceptual configuration is shown in Fig. \ref{Yb setup}. 
Here, the pulse  of approximately 200 fs obtained from the Yb:KGW amplifier is divided in two, i.e., one for the seed generation and another for the pump. 
On the seed generation side, the bandwidth from the Yb:KGW amplifier is broadened using the multi-pass plate method to generate a CEP-stabilized broadband seed via intra-pulse DFG. 
Note that the pulse for seed generation must be compressed to less than approximately 10 fs; thus, a two-stage multiplate sandwiching a chirped mirror is used \cite{MPlateC1,MplateC2,MPlateC3,MPlateC4} . 
The compressed pulse is focused on a GaSe crystal, and DFG is performed to prepare the CEP-stabilized broadband seed. 
The generated seed is split into two, passed through two kinds of AOPDF employed in parallel to control the seed dispersion, and combined by a beam combiner. 
The pump for the DC-OPA is amplified using a Yb-based TDL CPA.
 After the pulse stretcher, the 1 $\mu$m pulse for CPA is amplified by a thin-disk regenerative amplifier and multi-pass amplifier, and compressed by a grating pulse compressor. 
Typically, the obtained 1 $\mu$m pulse has a pulse width of approximately 1 ps and a spectral bandwidth of 5 nm. 
Here, a multi-pass cell method \cite{highpowerPC1,highpowerPC2} is employed to broaden the spectrum for DC-OPA.
Using this method, the spectral bandwidth of 1 $\mu$m pulse is expanded from 0.98 $\mu $m to 1.08 $\mu $m, and the dispersion is adjusted to the optimal chirp amount for the DC-OPA using chirped mirrors.
Then, DC-OPA is performed using the prepared pump and seed. 
When the seed energy is approximately 10 nJ, a three-stage DC-OPA is prepared to suppress the parametric fluorescence sufficiently. 
The expected performance is shown in Table \ref{Prospect of the DC-OPA}.
Note that the average power of the DC-OPA depends on the average power of the Yb-based TDL.
When a 200 W Yb-based TDL at a 10 kHz repetition rate is employed for the pump for the DC-OPA, the average power of a single-cycle laser will achieve 10 W at the central wavelength of 2.7 $\mu$m with an energy level of 1 mJ.

\FloatBarrier

Recently, high-average power Tm-doped fiber lasers near 1.9$\,\mu $m have been alsof developed, and it is becoming possible to develop high-average power picosecond lasers near 2$\,\mu $m using Ho-doped crystal pumped by a Tm-doped fiber laser. 
Here, we discuss a scheme to produce a single-cycle DC-OPA with a wavelength region of 3.5$\,\mu $m to 7$\,\mu $m pumped by Ho-doped TDL.

\begin{figure}
\centering\includegraphics[width=0.6\textwidth]{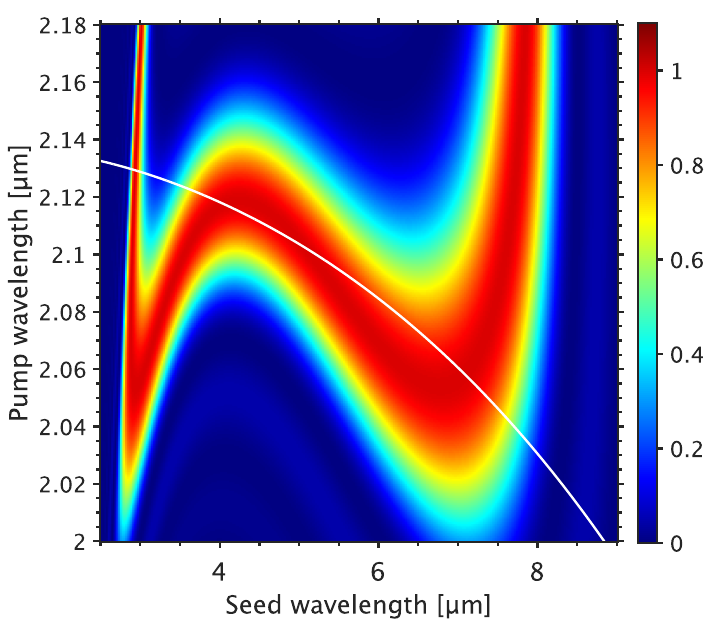}
\caption{PM and chirp matching to amplify a single-cycle 5$\,\mu $m laser pumped by Ho-based TDL with type I $\rm{LiInSe_2}$ $(\theta =23.5^{\circ}$).} \label{Ho PM}
\end{figure}

The PM and chirp matching condition of DC-OPA pumped by Ho-based TDL (2$\,\mu $m) for generating a high average power 5$\,\mu $m single cycle laser is shown Fig. \ref{Ho PM}. 
Here, the nonlinear crystal is a type I $\rm{LiInSe_2}$ $(\theta=23.5^{\circ})$ crystal. 
In Fig. \ref{Ho PM}, the white line shows the chirp matching between the pump and the seed. 
When a $\rm{CaF_2}$ bulk compressor with thickness of 40 mm was used and the pump dispersion is $\rm{2.2 \times 10^5\,fs^2}$, it is possible to amplify wavelength region of 3.5$\,\mu $m to 8$\,\mu $m simultaneously.  
The wavelength of the Ho-based TDL is approximately 2.09 $\mu $m, and, as with the Yb-based DC-OPA, this can be broadened to approximately 2.05$\,\mu$m to 2.14$\,\mu$m using a multi-pass cell. 
Note that the conceptual setup of the laser system is similar to the Yb-based DC-OPA system (Fig. \ref{Yb setup}). 
A compressed pulse from a multi-pass plate is employed as the front-end laser (Fig. \ref{Yb setup}) and split into two. 
Here, one beam is employed for DFG in BiBO to generate the 2 $\mu $m pump, and the other beam is employed in GaSe for the seed (3.5--7.0 $\mu $m) generation.
 On the pump generation side, the 2.09$\,\mu $m light from DFG for CPA is amplified by a Ho-based TDL CPA using a regenerative amplifier and a multi-pass amplifier with a Tm-doped fiber continuum wave laser as the pump for the Ho-doped TDL, and the amplified 2.09$\,\mu $m pulse is broadened by a multi-pass cell. 
Here, amplified 2.09$\,\mu $m pulse is adjusted to the optimal dispersion for the DC-OPA, and then used as the pump for the DC-OPA. 
As shown in Table \ref{Prospect of the DC-OPA}, the output MIR laser is expected to be 1.29 cycles.
When a 500 W Ho-based TDL with a 10 kHz repetition rate is used as the pump for the DC-OPA, the average power of a single-cycle laser achieves 10 W at the central wavelength of 4.6 $\mu$m with an energy level of 1 mJ.

\FloatBarrier

\section{Conclusion}
\label{Conclusion}
This paper has reviewed the development of MIR lasers using the DC-OPA method.
The DC-OPA method can generate more than 100 times the peak power compared with MIR sources using the conventional OPA technique. 
By optimizing the PM and chirp matching condition accurately, we have demonstrated multi-TW MIR lasers at several wavelength regions. 
Furthermore, we have demonstrated that the advanced DC-OPA method with heterogeneous nonlinear crystals enables one octave amplification bandwidth, which has previously been a difficult task, and we succeeded in developing a TW-class single-cycle laser. 
These DC-OPA lasers were applied to HHG, and we show the way to the application prospects in single-shot absorption spectroscopy, sub-GW IAP generation, and single cycle IAP generation.
As a future prospect of DC-OPA, this paper has discussed a path to the development of a TW-class sub-cycle DC-OPA pumped by a Ti:sapphire laser and high average power single-cycle DC-OPA pumped by a TDL.
The continuum region of HH driven by a sub-cycle pulse exceeds 70 $\%$; thus, it can be used to generate both a single-digit attosecond pulse width and a single-cycle soft X-ray pulse.
We expect that the development of MIR laser using the DC-OPA will facilitate further advances in ultrafast science and approach zepto-second pulse generation.
\section*{Acknowledegment}
 MEXT Quantum Leap Flagship Program (Q-LEAP) (JP-MXS0118068681); Ministry
of Education, Culture, Sports, Science and Technology (17H01067, 19H05628, 21H01850).
This project was partially supported by the RIKEN TRIP initiative (Leading-edge semiconductor technology).

\bibliography{DCOPA_review_ref2}

\end{document}